\pdfoutput=1
\documentclass[preprint,cha,amsmath,amssymb,aps,nofootinbib,showkeys,11pt]{revtex4-1}

\usepackage{graphicx}
\usepackage{dcolumn,mathtools}
\usepackage{bm}
\usepackage{pgfplots}
\pgfplotsset{width=10cm,compat=1.9}
\usepgfplotslibrary{fillbetween}
\newcommand{\ds}{\displaystyle}

\graphicspath{{Figures/}}

\begin{document}

\title{Bystander effect emerges from individual psychological prospects}

\author{Tiffanie Ng}
\email{ng2@kenyon.edu}
\affiliation{
Department of Mathematics and Statistics,
Department of Economics,
Kenyon College
}

\author{Sara M.~Clifton}
\email{cliftons@denison.edu}
\affiliation{
Department of Mathematics,
Denison University
}

\date{\today}

\begin{abstract}
The bystander effect is a social psychological phenomenon in which individuals are less likely to help a person potentially in need if there are others present. Sociologists and psychologists have proposed multiple plausible reasons for the bystander effect, from situational ambiguity and social contagion to diffusion of responsibility and mutual denial. We build a new model of an individual’s decision to intervene in a bystander situation based on these social psychological hypotheses, along with ideas borrowed from prospect theory. This model shows, for the first time, that the bystander effect emerges from social risk perception among non-coordinating individuals in ambiguous bystander situations. Expanding upon this static model, we explore the effect of social learning, where individuals update their perceived risk of intervening after experiencing or witnessing the social repercussions of previous interventions. A novel result of this model is that social learning exacerbates the bystander effect. We validate these models using a new database of 42 experimental and observational studies across a wide range of bystander situations, demonstrating a straightforward and generalizable explanation for the observed phenomenon, which may suggest effective interventions tailored to specific bystander situations.
\end{abstract}


\maketitle

\section{Introduction}
The bystander effect (also called \textit{bystander apathy} or \textit{bystander inhibition}) is a social phenomenon in which individuals are less likely to offer assistance to someone potentially in need as the number of witnesses to the situation grows \cite{BystanderApathy}. Research of the bystander effect was pioneered by social psychologists John M.~Darley and Bibb Latan{\'e} who, in the 1960s, published a series of papers examining the various situational and psychological factors from which the effect emerges \cite{latane1968whenpplhelp, latane1968inhibition, darley1968diffusion, latane1969apathy}. 

Early theories of the social psychological mechanisms underlying the bystander effect include diffusion of responsibility (a rational division of obligation among all witnesses), audience inhibition (a fear of public embarrassment in a large crowd), and social influence (a hesitancy to act unless others are also acting) \cite{fischer2011metaanalysis}. However, the mechanism(s) likely depends on the type of bystander situation under consideration. Bystander situations are broadly categorized by the degree and type of danger, if any; by the degree of situational ambiguity; and by the degree of witness competence to assist \cite{fischer2011metaanalysis}. 

Within those broader categories, studies have explored the impact of victim/witness demographics (e.g., age \cite{staub1970child,plotner2015young,bauman2020experiences}, race \cite{gaertner1982race,york2016racial,garcia2022racial}, gender \cite{latane1975sex,cox2018bystander,tice1985masculinity,jenkins2017bullying}), environmental factors (e.g., rural versus urban \cite{merrens1973nonemergency}, lab versus field \cite{shaffer1975intervention}), and social or physical connections among witnesses or victims (e.g., friends versus strangers \cite{latane1969lady,seo2022helping,bennett2016friends}, able to communicate versus separated \cite{latane1981ten,krueger2009rational}, in person versus online \cite{you2019bystander,voelpel2008david,blair2005electronic,voelpel2008david,markey2000bystander}). Metrics of interest have included not only the intervention rate, but also latency of reaction \cite{schwartz1976theft,darley1968diffusion} and the extent of action taken \cite{barron2002email,harari1985rape}.

Although the bystander effect is a robust phenomenon observed in a variety of situations \cite{fischer2011metaanalysis}, the effect is not universal. In fact, the bystander effect is reduced or reversed when the situation is clearly dangerous and the witnesses are competent to assist \cite{fischer2006unresponsive,franzen2013volunteer}. A bystander effect is most commonly observed when the situation is ambiguous \cite{clark1974apathetic,solomon1978helping} and not physically dangerous, suggesting that the phenomenon may be primarily driven by social risk rather than physical risk \cite{karakashian2006fear, siligato2024freezing}. Therefore, we focus our study on ambiguous, non-dangerous bystander situations in which witnesses are unsure what the appropriate behavior is, and witnesses might not agree on the appropriateness of action. 

\subsection{Background}
Many qualitative models for the bystander effect have been developed and tested, including the five-step Bystander Intervention Model \cite{latane1970unresponsive}, in which a witness must (1) notice the event, (2) interpret that a person might need help, (3) decide on personal responsibility, (4) decide if they can help, and (5) consciously decide to act or not. A bystander effect can occur due to decisions at any stage; specifically, situational ambiguity comes into play in Step 2, diffusion of responsibility or social influence in Step 3, witness competence in Step 4, and audience inhibition in Step 5. Building on the Bystander Intervention Model, Piliavin, Dovidio, Gaertner, \& Clark proposed the Arousal Cost-Reward Model in 1981 \cite{piliavin1982cooperation,pilavin1991arousalcostreward}, positing that people experience discomfort when seeing someone in need, and they attempt to efficiently reduce their discomfort with the highest reward and/or the lowest cost. The considered costs and rewards of acting or not acting are primarily emotional (e.g., pride, embarrassment, guilt, fear) rather than rational (e.g., money, time, safety), and decisions may not be conscious.

Quantitative models of the bystander effect are quite limited, but the few that exist utilize a variety of mathematical frameworks spanning a wide range of complexity; the models typically incorporate a decision process consistent with the Bystander Intervention Model or the Arousal Cost-Reward Model. Among the more complex models, a temporal-causal network model of an individual's decision state qualitatively aligns with expected arousal level dynamics \cite{van2020help}. Among the simpler modeling approaches, a game-theoretic model shows that the rational costs/benefits of acting or not acting produce a bystander effect, justifying smaller class sizes to reduce bullying \cite{isada2016economic}. The simplest mathematical model of a bystander effect shows that gratuity rates at a steakhouse in Columbus, Ohio, follow a power law with power $\approx 0.22$ \cite{freeman1975diffusion}, attributed to diffusion of responsibility. Quantitative models of collective decision-making are more common (e.g., \cite{banerjee1992simple}), although they do not explicitly address the bystander effect. 

Because bystander situations often require quick (and potentially socially risky) decisions under uncertainty, we build a model that incorporates concepts from behavioral economics \cite{frank2010microeconomics}. In particular, prospect theory diverges from traditional economic frameworks by acknowledging that people do not make purely rational decisions when faced with uncertainty and the potential for gain or loss \cite{kahneman1979prospect}. Prospect theory is based on several generalizable observations: (1) people make decisions relative to the status quo (i.e., no action taken), (2) a loss is more painful than an equal gain (a concept called ``loss aversion''), and (3) people tend to overweight small probabilities and underweight large probabilities. Prospect theory acknowledges these observed irrational behaviors that break symmetries (e.g., between acting or not acting, or between gain and loss) assumed by  more traditional expected utility theory.

\section{Methods}
With the goal of understanding the plausible origin of the bystander effect, we build a model of individual decisions to intervene (or not) in a socially ambiguous, non-dangerous bystander situation based on the social costs and benefits of intervening (Section \ref{sec:bystander}). In the simple version of the model, we assume that all individuals perceive intervention as equally appropriate, and these views are fixed. We explore this model in Section \ref{sec:staticbehavior}. We then build an extension of the static bystander model by assuming that individuals learn from the outcomes of bystander situations (Section \ref{sec:learning}), discuss various frameworks to simulate the dynamics (Section \ref{sec:structure}), build and explore a simple all-to-all network framework (Section \ref{sec:dynamicbehavior}), and test the sensitivity of model outcomes to parameters (Section \ref{sec:sensitivity}). Finally, we describe a new database that we have compiled of observational and experimental bystander effects (Section \ref{sec:data}), against which we compare our model behavior.

\subsection{Bystander Model} \label{sec:bystander}
Imagine a social situation in which someone appears to require assistance, like a potentially elderly or pregnant person who may want a seat on a crowded bus. In such instances, witnesses in the situation have the choice of whether or not to take action.\footnote{In our approach, ``taking action'' implies that a person decided to act, whether or not the witness needed to follow through on the action. As an illustration, suppose that three people decide to give up their bus seats, but only the first person to stand up relinquishes their seat. In this case, we say that three people took action.} Making the decision to act in a socially ambiguous situation like this comes with a psychological risk. If the action is perceived as appropriate, they experience the emotional benefit of pride and righteousness. However, if their action is seen as inappropriate, they may experience embarrassment or shame. On the other hand, choosing not to act may feel closer to neutral. 

The psychological risk of taking action can be viewed as a prospect (a psychological gamble), with the default option being no action \cite{kahneman1979prospect}. Although choosing not to act provides no net psychological benefit, deciding to act involves the risk that the action will be received either positively (e.g., with appreciation or a nod) or negatively (e.g., with a frown or a condemnation). Each person who decides to act assesses the risk that someone else (a witness or a person in need of assistance) is dissatisfied with the action, which we represent by the value $r \in [0,1]$. The perceived probability that at least one witness will display dissatisfaction is calculated as $\displaystyle s = 1 - (1-r)^N$, where $N$ is the number of witnesses\footnote{This formulation assumes that the situation involves an individual potentially in need of assistance who may condemn action, but is not a witness as they cannot intervene to help themselves. To accommodate a situation with no such individual (e.g., smoke seeping into a room \cite{latane1968inhibition}), $\displaystyle s = 1 - (1-r)^{N-1}$. To accommodate a situation with more than one such person (e.g., assault or bullying \cite{fischer2006unresponsive, harari1985rape, gueguen2015commitment,kazerooni2018cyberbullying,shaffer1975intervention}), shift $N$ by the number of additional people involved who could in theory condemn an individual's intervention. Because this variety of bystander situations only shifts $N$, we only explore the most common scenario with one non-witness.}; here we assume that the probabilities of displaying dissatisfaction are independent. We then calculate the expected value of acting as $\displaystyle V=B(1-s)-Hs$ for a given witness, where $B$ and $H$ represent the psychological benefit or harm of an action, respectively. To reduce the number of parameters (and because $B$ and $H$ don't have accepted units), we divide the expected value equation by $B$ to obtain $v=1-s-sx$, where $x=H/B$ is a loss aversion ratio. Following principles of prospect theory \cite{kahneman1979prospect}, if the expected value of acting is greater than the value of not acting ($v>0$), the witness will choose to intervene; otherwise, they will not act.

In principle, each witness (indexed by $i$) may perceive a different risk value $r_i$. However, we first explore the case in which witnesses have homogeneous risk assessments; this is a special case in which all members of society agree on the appropriateness of intervention. Additionally, witnesses may have different loss aversion ratios $x_i$, and the average loss aversion ratio is known to typically fall between $1.5$ and $2.5$ \cite{arora2015risk, blake2021quantifying, kahneman2011thinking}.\footnote{These studies involve risk of gaining or losing money. Although it is qualitatively known that intervention propensity varies in a population \cite{hortensius2018empathy}, it is unknown if psychological risk is qualitatively or quantitatively similar to monetary risk.} 

As we expect population-level loss aversions to be approximately bell-curved while not permitting negative loss aversion, we suppose the population distribution of loss aversion ratios follows a gamma distribution\footnote{Results are similar for a log-normal distribution or a normal distribution with negligible negative left tail.}: 
\[x_i \sim \Gamma\left(\alpha = \left(\frac{\mu}{\sigma}\right)^2,\beta = \frac{\sigma^2}{\mu}\right),\] 
where $\mu$ and $\sigma$ are the mean and standard deviation of the gamma distribution, respectively. Given the probability density function (PDF) of loss aversion ratios is 
\[f_X(x) = \frac{x^{\alpha-1} e^{-x/\beta}}{\beta^\alpha \Gamma(\alpha)},\,\,\,\,\, x>0\]
the PDF of expected psychological values of acting is $\ds f_V(v) = \frac{1}{s} f_X\left( \frac{1-s-v}{s} \right)$. 

\begin{figure}[th]
\center{\includegraphics[width=0.8\textwidth]{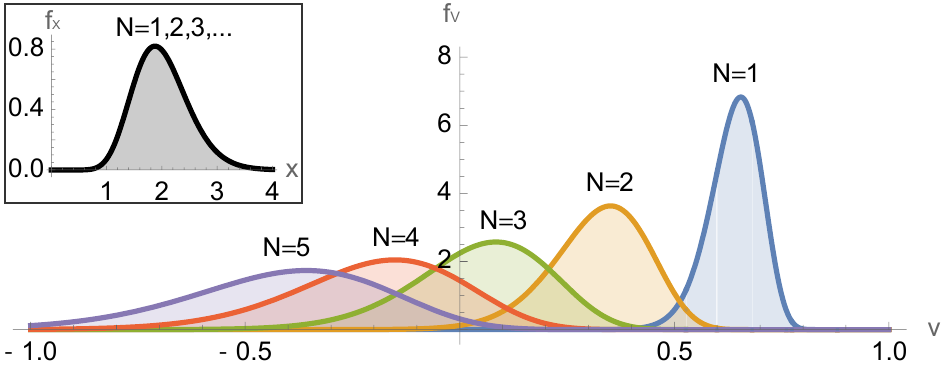} \\\includegraphics[width=0.7\textwidth]{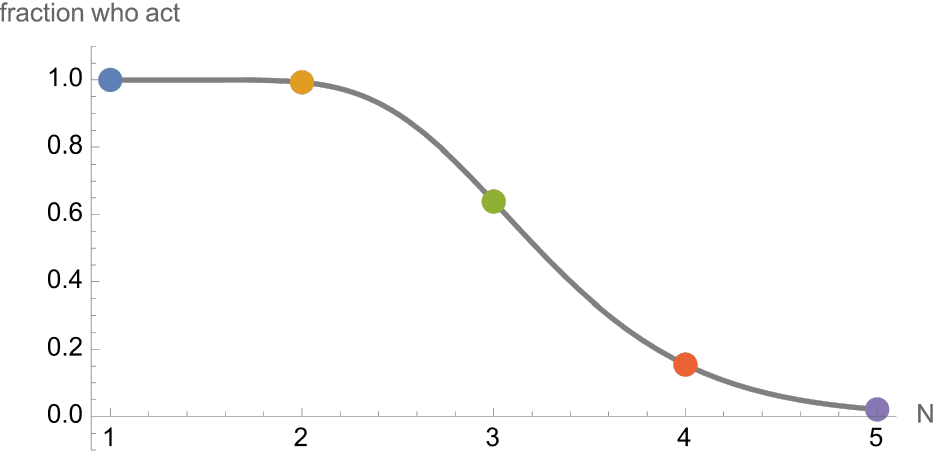}}
\caption{Assuming a gamma distribution of loss aversion ratios with $\mu=2$ and $\sigma=0.5$ (top inset) following experimentally determined values \cite{arora2015risk,blake2021quantifying,kahneman2011thinking}, the corresponding distributions of expected values of acting shift left as crowd size $N$ increases (top). For this example, $r=0.12$ is the perceived chance that a particular person will condemn intervention in a bystander situation. The fraction of $f_V$ that falls in the range $v\in(0,\infty)$ represents the proportion of individuals who would intervene ($I$), which decreases as the crowd size grows (bottom).}
\label{fig:dist}
\end{figure}

In a given scenario with $N$ witnesses, the expected fraction of witnesses who will act can be determined by finding the fraction of the value distribution that is positive: 
\begin{align}
    I = \int_0^{\infty} f_V(v) \, \mathrm{d}v = 1 - \frac{y^{\alpha} E_{1-\alpha}(y)}{\Gamma(\alpha)} = 1 - \frac{\Gamma(\alpha,y)}{\Gamma(\alpha)}, \label{eq:static}
\end{align}
where $\ds y = \frac{(1-r)^N}{\beta \big(1-(1-r)^N\big)}$, $E_a(x)$ is the exponential integral, and $\Gamma(a,x)$ is the incomplete gamma function. We see that as $N$ increases, the fraction who intervene decreases; this is an emergent bystander effect, which occurs even if all witnesses agree on the appropriateness of an action (Figure \ref{fig:dist}).
 
\begin{figure}[th]
\center{\includegraphics[width=0.6\textwidth]{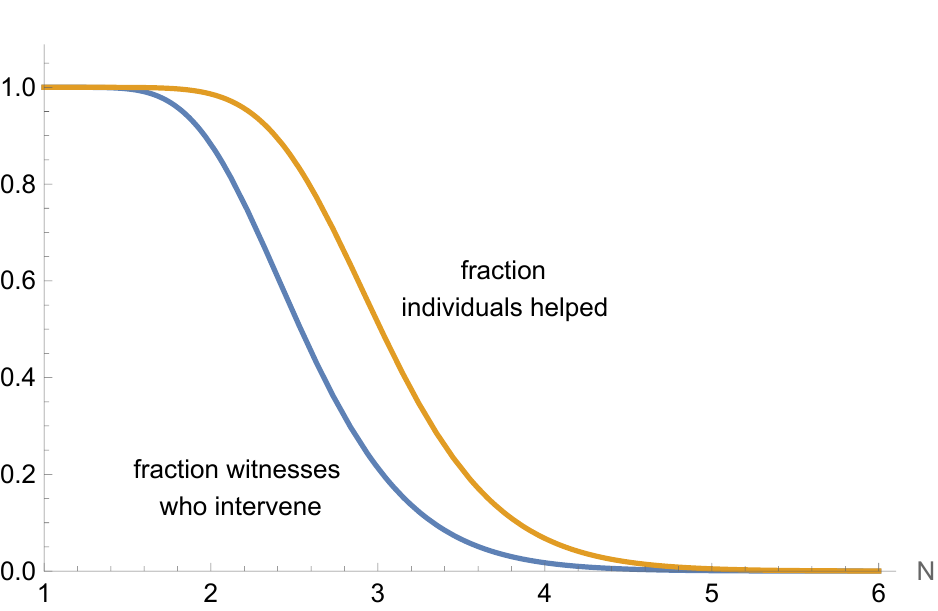}}
\caption{Relationship between the fraction of witnesses who will intervene in a bystander situation versus the fraction of victims in a bystander situation that will be helped, shown here with a gamma distribution of loss aversion ratios with $\mu=2$ and $\sigma=0.5$ and with action appropriateness $r=0.15$. The bystander curves retain the same qualitative shape, but are shifted right (for any given number of witnesses, the probability the witness is helped is larger than the probability that any given witness helps).}
\label{fig:help}
\end{figure}

Although many sociological experiments record the fraction of individuals who intervene in a bystander situation (our metric of interest), other experiments only record the fraction of bystander situations in which the victim received assistance from at least one witness. To convert to this metric, we assume the probability of intervening for an individual is $p$; then the probability that at least one witness intervenes (the victim is helped) is $\ds 1-(1-p)^N$, the same formula used by Latan{\'e} and Nida \cite{latane1981ten}. Both curves are plotted with $\ds r=0.15, \mu=2, \sigma=0.5$ in Figure \ref{fig:help}. We see that the relationships between crowd size and fraction intervene/fraction victims helped are similar. 

\subsection{Static Model Behavior} \label{sec:staticbehavior}
\begin{figure}[th]
\center{\includegraphics[width=0.49\textwidth]{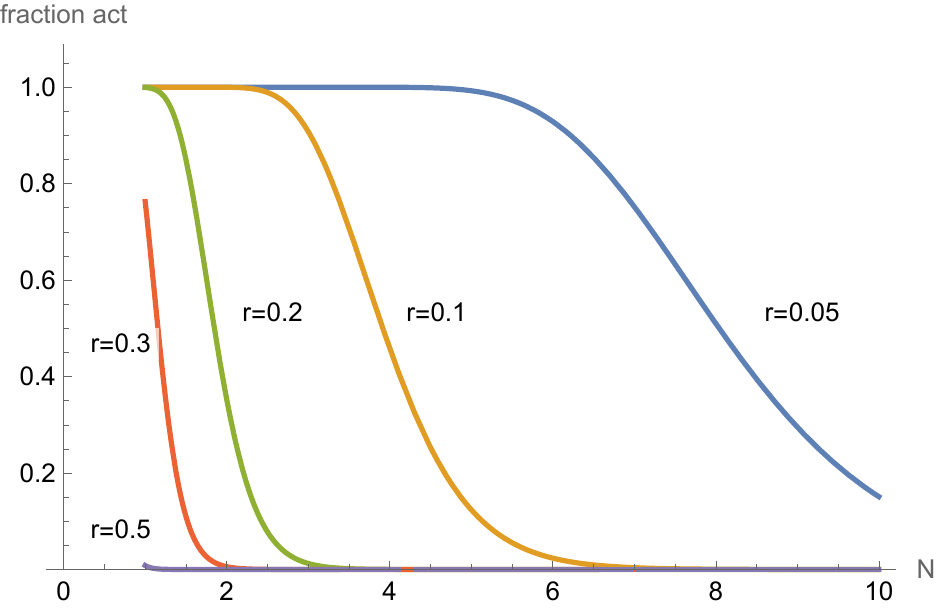} \\ \includegraphics[width=0.49\textwidth]{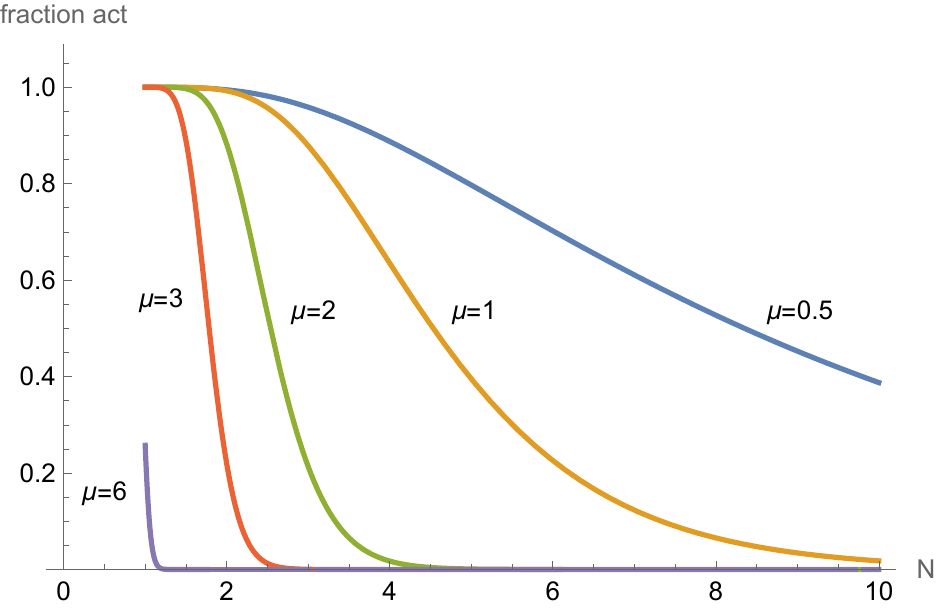}
\includegraphics[width=0.49\textwidth]{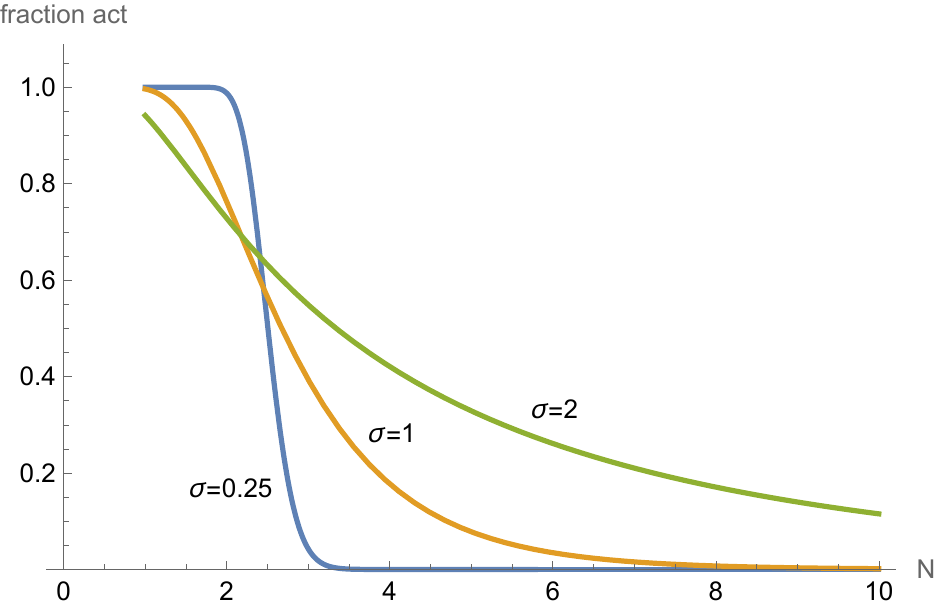}}
\caption{Fraction of bystanders who intervene as a function of the number of bystanders $N$. Unless otherwise stated, we assume a gamma distribution of loss aversion ratios with $\mu=2$ and $\sigma=0.5$ \cite{arora2015risk, blake2021quantifying,kahneman2011thinking} and action appropriateness $r=0.15$. Holding all else constant, increasing action appropriateness $r$ causes the bystander curve to shift left (top center). Increasing the average loss aversion ratio causes the bystander curve to compress left (bottom left). Increasing the loss aversion standard deviation causes the large-$N$ intervention fraction to increase, while the small-$N$ intervention fraction decreases (bottom right).}
\label{fig:param_SA}
\end{figure}

Under the assumption that appropriateness of intervening is constant and homogeneous, we explore how the model behavior changes as we vary action risk $r$, and the mean $\mu$ and standard deviation $\sigma$ of the loss aversion ratio distribution (see Figure \ref{fig:param_SA}). Holding all else constant, increasing action risk $r$ causes the bystander curve to shift left. This implies that as intervention feels more appropriate, bystanders are more willing to act for any given crowd size (e.g., socially appropriate actions such as holding the door for a person carrying groceries are more likely, while socially risky actions like touching someone without their permission are less likely).

Again holding all else constant, increasing the average loss aversion ratio causes the bystander curve to compress left. This implies that as public embarrassment becomes more painful, bystanders are less likely to intervene in any given situation. 

More surprisingly, increasing the loss aversion standard deviation causes the large-$N$ intervention fraction to increase, while the small-$N$ intervention fraction decreases. This implies that as the population becomes less homogeneous in their loss aversion, the bystander effect becomes less extreme (i.e., the fraction of individuals willing to intervene drops more slowly as crowd size grows).  

\subsection{Social Learning Model} \label{sec:learning}
Because appropriateness of intervening is learned through repeated exposure to bystander situations, we next explore how individuals' perceived risk $r_i$ evolves as they learn from their own actions and the actions of others. Suppose that witnesses update their risk $r_i$ under four scenarios: the witness intervened and it was received either poorly or well, or the witness did not intervene, but someone else did and it was received either poorly or well. If actions by the individual or others are received poorly, then the witnesses will increase their risk of action, while well-received actions will decrease the risk of action. We incorporate these four learning mechanisms by assuming that each individual's risk is updated continuously according to the differential equation
\begin{align}
	\frac{\mathrm{d}r_i}{\mathrm{d}t} = \underbrace{r_i (1-r_i)}_{\text{restrict } r\in[0,1]} 
 \bigg( &\underbrace{a \, P_i^{(I)} (1-P_i^{(w)})}_{\text{act received poorly}}
 - \underbrace{b \, P_i^{(I)} P_i^{(w)}}_{\text{act received well}} 
 + \underbrace{c \, (1-P_i^{(I)}) \tilde{P}_i^{(p)}}_{\text{others' act received poorly}}
 - \underbrace{d \, (1-P_i^{(I)}) \tilde{P}_i^{(w)}}_{\text{others' act received well}}  \bigg), \label{eq:odes}
\end{align}
where $a,b,c,d$ are the learning rates for the four scenarios above, and $\ds P_i^{(I)}$, $\ds P_i^{(w)}$, $\ds \tilde{P}_i^{(p)}$ and $\ds \tilde{P}_i^{(w)}$ are the probabilities of individual $i$ intervening, individual $i$'s action being received well, and another individual's action being received poorly or well, respectively, as described below.

The probability that individual $i$ intervenes is 
\begin{align*} 
P_i^{(I)} &= \begin{dcases}
                    1, & 1-(1+x_i)s_i > 0 \\
                    0, & 1-(1+x_i)s_i \le 0
                \end{dcases}
                = H(1-(1+x_i)s_i)
\end{align*}
where $H$ is the Heaviside function, and  $\ds s_i = 1 - (1-r_i)^N$, capturing that each individual assumes everyone else holds a risk level equal to their own (no one is a mind-reader).

The probability that individual $i$'s action is well-received, given that the individual acted, is
\begin{align*}
    P_i^{(w)} = \prod_{j\ne i} \bigg( P_j^{(I)} + \big(1-P_j^{(I)}\big) (1-r_j) \bigg)
\end{align*}
We derive this probability in the Supplemental Materials.

The probability that another witness's action is well-received, given that individual $i$ did not intervene, is
\begin{align*}
    \tilde{P}_i^{(w)} = \prod_{j\ne i} \bigg( P_j^{(I)} + \big(1-P_j^{(I)}\big) (1-r_j) \bigg) - \prod_{j\ne i} \big(1-P_j^{(I)}\big) (1-r_j),
\end{align*}
where the first product captures all possible outcomes in which no one is ``called out'' for their behavior, and the second product corrects for the outcome in which no one acts and therefore individual $i$ cannot learn from the situation.

Finally, the probability that another witness's action is poorly-received, given that individual $i$ did not intervene, is
\begin{align*}
    \tilde{P}_i^{(p)} = 1 - \tilde{P}_i^{(w)} - \prod_{j\ne i} \big(1-P_j^{(I)}\big),
\end{align*}
where the product corrects for the outcome in which no one acts and therefore individual $i$ cannot learn from the situation. See Supplemental Materials for a visual derivation of these probabilities.

	\begin{table}[ht]
	\caption{Description of model parameters in social learning model \eqref{eq:odes}.}
\footnotesize
	\begin{tabular}{| c  p{8.5cm} p{1.8cm}  p{1.5cm} p{2cm} |}  \hline 
	{\bf Parameter} & {\bf Meaning} & {\bf Range} & {\bf Baseline} & {\bf Sources} \\  \hline 
    $N$ & number of witnesses to an individual potentially in need of help$^{\dag}$ & [1, 20] & variable & --- \\ 
	$a$ & risk update rate when learning from one's own poorly received actions & 0 or 1 & 1$^{\ddag}$ & --- \\ 
	$b$ & risk update rate when learning from one's own well received actions & [0, 100] & 0.01$^{\S}$ & \cite{yin2023differential,wachter2009differential,sidowski1956influence,gregory2007effects,jones2021increased} \\ 
	$c$ & risk update rate when learning from others' poorly received actions & [0, 100] & 0.1$^{\star}$ & \cite{moraes2025unravelling,riley2017active,blanie2018impact,peruch2004active} \\ 
    $d$ & risk update rate when learning from others' well received actions & [0, 100] & 0.001$^{\S}$ & \cite{yin2023differential,wachter2009differential,sidowski1956influence,gregory2007effects,jones2021increased} \\ 
	$x_i$ & loss aversion ratio for individual $i$, sampled from the gamma distribution $\displaystyle x_i \sim \Gamma\left(\alpha = \left(\frac{\mu}{\sigma}\right)^2,\beta = \frac{\sigma^2}{\mu}\right)$ & [0,100]  & variable & \cite{kahneman1979prospect} \\
    $\mu$ & mean of loss aversion ratio distribution & [0, 100] & 2 & \cite{arora2015risk, blake2021quantifying,kahneman2011thinking} \\
    $\sigma$ & standard deviation of loss aversion ratio distribution & [0, 100] & 0.5 & \cite{arora2015risk, blake2021quantifying, kahneman2011thinking} \\
    $r_{\text{max}}$ & maximum initial risk level, where initial risk is sampled from the uniform distribution $r_i(0) \sim \mathcal{U}(0,r_{\text{max}})$ & [0,1] & 0.1 & --- \\
	\hline
	\end{tabular}
	 \begin{flushleft} \footnotesize
  $^{\dag}$ In our model, we have assumed that there is a single individual capable of condemning intervention (e.g., not an animal or infant). If the bystander situation has no victim (e.g., smoke is seeping into a room), then subtract 1 from $N$. If the situation has more than one non-witness (e.g., bullying or a domestic argument), add the additional non-witnesses to $N$. \\
	 $^{\ddag}$ without loss of generality, rate may be assigned to 1, setting the timescale of the model \\
	 $^{\S}$ Research on learning rates from positive versus negative feedback is mixed. In physical tasks with physical reward and punishment, many studies show that learning is quicker (but potentially less generalizable) with negative reinforcement \cite{yin2023differential,wachter2009differential,sidowski1956influence}. In studies involving verbal feedback, it is also widely observed that people learn more quickly from feedback on their errors than on their correct responses to factual questions \cite{gregory2007effects}, but the strength of this effect may depends on age and gender \cite{eppinger2011choose,van2008evaluating}. Among studies that consider the emotions stimulated by feedback, anxiety is the most commonly measured; some studies show that anxiety enhances learning from negative feedback \cite{jones2021increased}, while others show the opposite \cite{petzold2010stress}. To our knowledge, no studies quantify the relative rate of learning from positive versus negative feedback in the types of social situations we are trying to understand, so we guess that learning is 100 times slower after positive feedback (especially since positive feedback could simply be the absence of negative feedback), and we test the sensitivity of model results to this choice in Section \ref{sec:sensitivity}.  \\
  $^{\star}$ Research on learning rates from `doing' versus `observing' is not definitive. Many studies show performance (particularly in physical tasks) is stronger after actively learning \cite{moraes2025unravelling,riley2017active,blanie2018impact,peruch2004active}, while other studies show no significant difference in performance (or occasionally the opposite effect) in an active versus passive learning environment \cite{reime2017learning,haidet2004controlled,chalmers1974learning}. To our knowledge, no studies measure learning rates (only final performance), nor do they measure the relative rates of learning from personal embarrassment versus empathetic embarrassment \cite{miller1987empathic,hawk2011taking} in the types of social situations we are trying to understand. Therefore, we guess that learning is 10 times slower from observed actions than from performed actions, and we test the sensitivity of model results to this choice in Section \ref{sec:sensitivity}.
	  \end{flushleft}
	\label{tab:param}
	\end{table}

\subsection{Dynamic Model Structures} \label{sec:structure}
There are at least three reasonable ways to simulate the model dynamics. The simplest strategy assumes that the population consists of $N$ individuals that continuously interact (i.e., individuals are connected on an all-to-all social network); this leads to a system of $N$ ordinary differential equations prescribed by Equations \eqref{eq:odes}. We focus on this simple option in this paper, and we explore other options in the Supplemental Materials.

More realistically, one might assume that a large finite population interacts on a randomly chosen (but static) network; this leads to a dynamic network model with a node representing each population member. We explore this option in Section \ref{sec:model2}. One might also assume that the network structure itself is dynamic; this analysis is beyond the scope of this work.

Finally, one may assume $N$ is a parameter within a continuum population, which leads to a single partial differential equation for the population distribution of risk $r$. We derive this partial differential equation in Section \ref{sec:model3}, but we leave its analysis for future work.

\subsection{Dynamic Model Behavior} \label{sec:dynamicbehavior} 
If $N$ individuals continuously learn by experiencing bystander situations with $N$ witnesses, the system can be viewed as a dynamic all-to-all network (i.e., a complete graph). We begin exploring model behavior under the simplifying assumption that learning from others' behavior is negligible relative to learning from one's own actions ($c=d=0$)\footnote{It is not reasonable to assume that learning by observing is negligible \cite{reime2017learning,haidet2004controlled,chalmers1974learning}. We explore this case simply to test the role of learning from doing before we add in the role of learning from observing.}. 

\begin{figure}[th]
\includegraphics[width=0.99\textwidth]{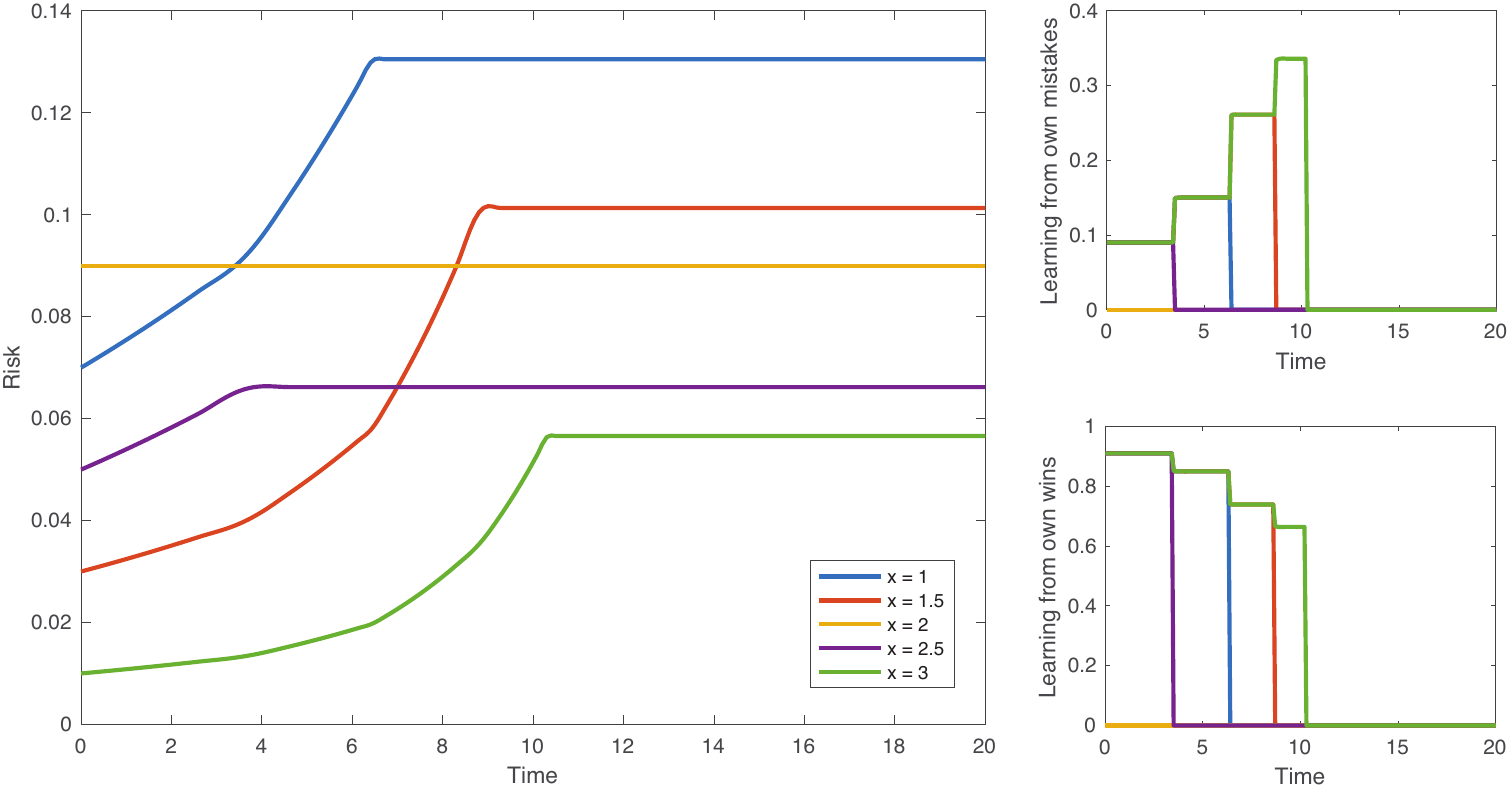}
\caption{Representative simulation sampling initial risks from $r_i(0) \sim \mathcal{U}(0,0.1)$ with $N=5$ and only learning from one's own actions: $a=1, b=0.01, c=d=0$ (left).
Contributions to social learning from one's own poorly-received actions (top right) and one's own well-received actions (bottom right), as measured by the probability of those outcomes occurring.
We see that learning is piecewise in time, with learning only occurring while the individual is taking action. For instance, `yellow' begins with a high risk and never learns because they never act, while `green' begins with a low risk and learns from a gradually increasing probability of condemnation until they finally stop acting.}
\label{fig:behav1}
\end{figure}

In a typical simulation (see Figure \ref{fig:behav1}), social learning from one's own actions increases personal risk $r_i$ until the individual ceases to act and therefore ceases to learn. Individuals do not reach their steady risk level simultaneously, nor do they cease acting because the probability of positive or negative reactions to their actions begin to balance; instead the probability of acting suddenly drops to zero for an individual as the value of acting passes the zero threshold (due to gradually increasing risk). This behavior is common among piecewise smooth dynamical systems \cite{bernardo2008piecewise}.

For different random initial conditions, it is also possible that all individual risks exponentially decay to zero. This occurs when all individuals begin with a low enough risk that they all initially act, and the probability of condemnation from all others remains zero. In fact the eventual steady state of the system is quite sensitive to initial conditions (see Sensitivity Analysis, Section \ref{sec:sensitivity}).  

\begin{figure}[th]
\includegraphics[width=0.99\textwidth]{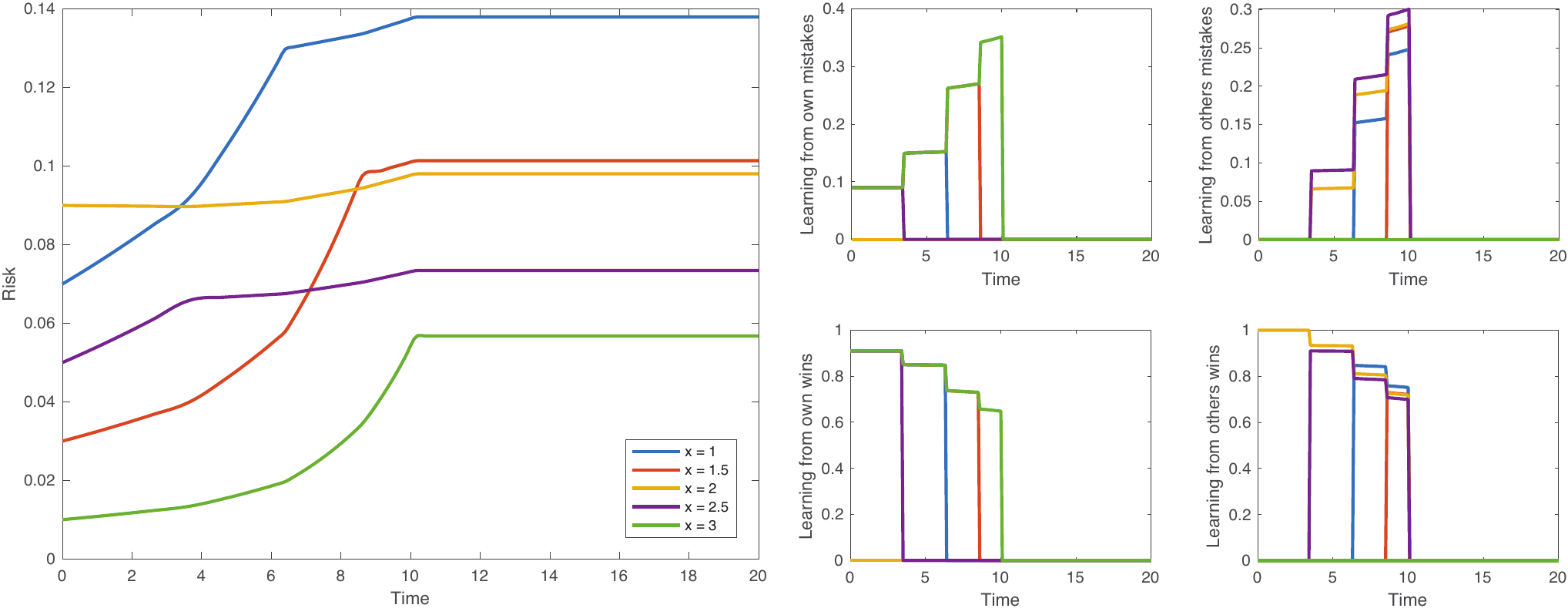}
\caption{Representative simulation sampling initial risks from $r_i(0) \sim \mathcal{U}(0,0.1)$ with $N=5$ and learning from reactions to self and others' actions: $a=1, b=0.01, c=0.1, d=0.001$ (left), based on evidence that learning from experience is faster than learning from observation \cite{moraes2025unravelling,riley2017active,blanie2018impact,peruch2004active} and learning from negative outcomes is faster than learning from positive outcomes \cite{yin2023differential,wachter2009differential,sidowski1956influence,gregory2007effects,jones2021increased}. 
Contributions to social learning from one's own poorly-received actions (top middle), one's own well-received actions (bottom middle), others' poorly-received actions (top right), others' well-received actions (bottom right), as measured by the probability of those outcomes occurring.
When individuals also learn from others' actions, we see that witnesses switch from learning exclusively from their own actions to learning from others' actions. For instance, `blue' begins by learning from their own actions until their learned risk reaches a threshold when they stop acting; after that time, `blue' continues to learn increased risk by watching others get condemned for acting.}
\label{fig:behav2}
\end{figure}

One unrealistic outcome of the simplifying assumption that individuals only learn from their own actions is that those who don't act cannot update their risk level. 
Under the more realistic assumption that individuals learn from both their own and from others' behavior, witnesses switch from learning exclusively from their own actions to learning from others' actions. Because the probability of acting is either 0 or 1, witnesses cannot learn simultaneously from acting and not acting. In a typical simulation (see Figure \ref{fig:behav2}), witnesses begin by learning from their own actions until their learned risk reaches a threshold at which they stop acting; after each witness stops acting, they continue to learn increased risk by watching others' get condemned for acting.

\begin{figure}[th]
\includegraphics[width=0.99\textwidth]{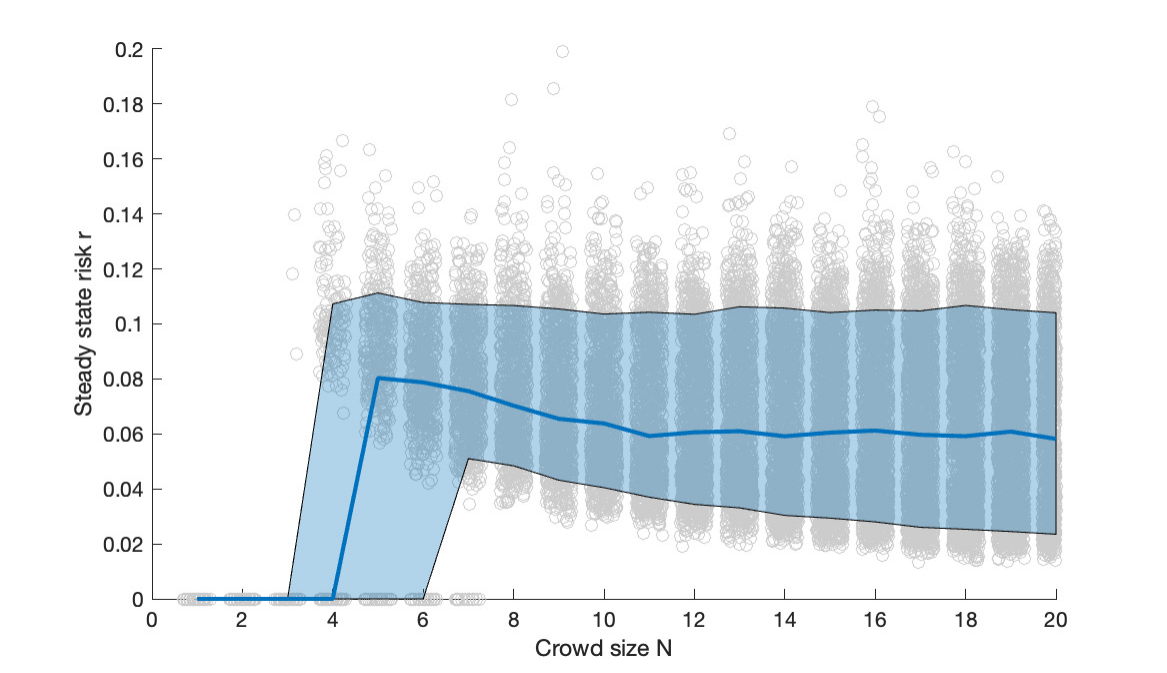}
\caption{Steady state learned risk on an all-to-all network as a function of the number of witnesses $N$, holding all else constant at the baseline values in Table \ref{tab:param}. Grey dots are individual witness steady state risk levels $r_i$ over 100 model simulations, jittered horizontally for easier visualization. The median steady state risk level for each $N$ (blue line) changes non-monotonically as $N$ increases: for small enough crowd size, all witnesses converge to zero risk; after a threshold crowd size, learned risk jumps above the median initial risk; as crowd size increases beyond the threshold level, risk decreases slightly and then remains approximately constant. The variance in learned risk tends to increase with $N$ (10\% to 90\% quantiles of simulation results for each $N$ shown with blue shading).}
\label{fig:behav3}
\end{figure}

Holding all else constant, the steady state learned risk level $r^*_i$ depends non-monotonically on the number of witnesses $N$ (see Figure \ref{fig:behav3}). Under the parameters given in Table \ref{tab:param}, all witnesses converge to zero risk for small enough crowd size. At a threshold crowd size (for our parameters values, around 3 to 6 witnesses), the median steady state risk jumps above the median initial risk. As crowd size increases beyond the threshold level, the median steady state risk decreases slightly and remains approximately constant beyond a crowd size of 10 to 12 (this is consistent with psychological research demonstrating that humans are not capable of quickly perceiving quantities larger than a handful \cite{trick1994small}).

\begin{figure}[th]
\includegraphics[width=0.8\textwidth]{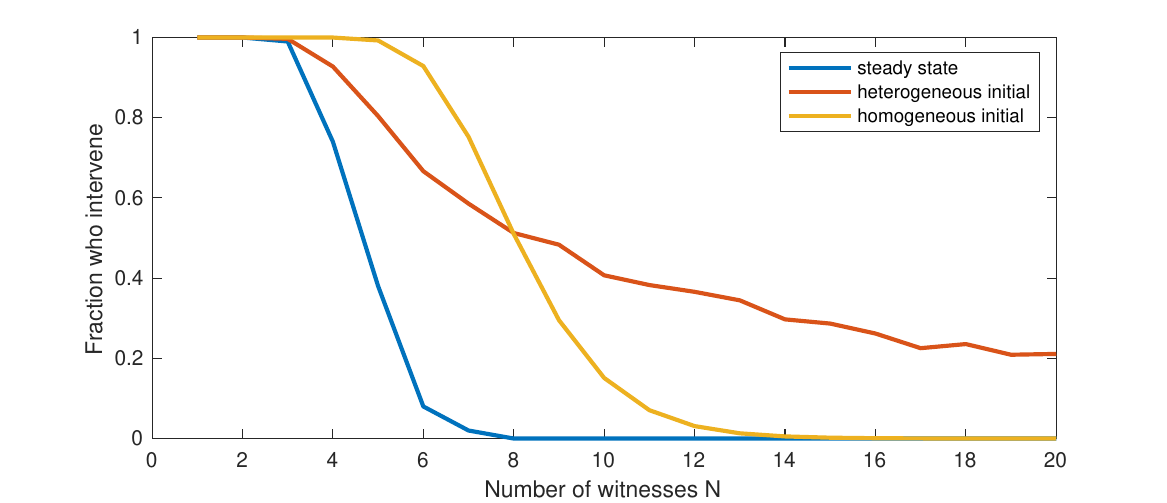}
\caption{Initial versus steady state intervention fraction as a function of the number of witnesses $N$, holding all else constant at the baseline values in Table \ref{tab:param}. After social learning (Eqn (\ref{eq:odes})), the bystander curve shifts left from the initial state (red) to the steady state (blue), implying that the bystander effect is exacerbated by social learning. We also show the initial bystander curve assuming that all bystanders initially hold identical risk perceptions equal to the average initial risk (yellow), illustrating that the assumption of homogeneous risk is not a good approximation for heterogeneous risk.}
\label{fig:behav4}
\end{figure}

\subsection{Sensitivity Analysis} \label{sec:sensitivity}
Because bystander situations vary widely in their social and cultural context, we conduct a sensitivity and uncertainty analysis to understand the impact of each model parameter on the observed outcome of interest: steady state perceived risk of action (we use the median population risk). Because for any given crowd size $N$ the relationships between the parameters and the median steady state risk levels are generally nonlinear but monotonic, we use Latin Hypercube Sampling (LHS) with $n=1000$ samples of parameter space and partial rank correlation coefficients (PRCCs) \cite{marino2008methodology}.  

\begin{figure}[th]
  \centering
    \includegraphics[width=0.99\textwidth]{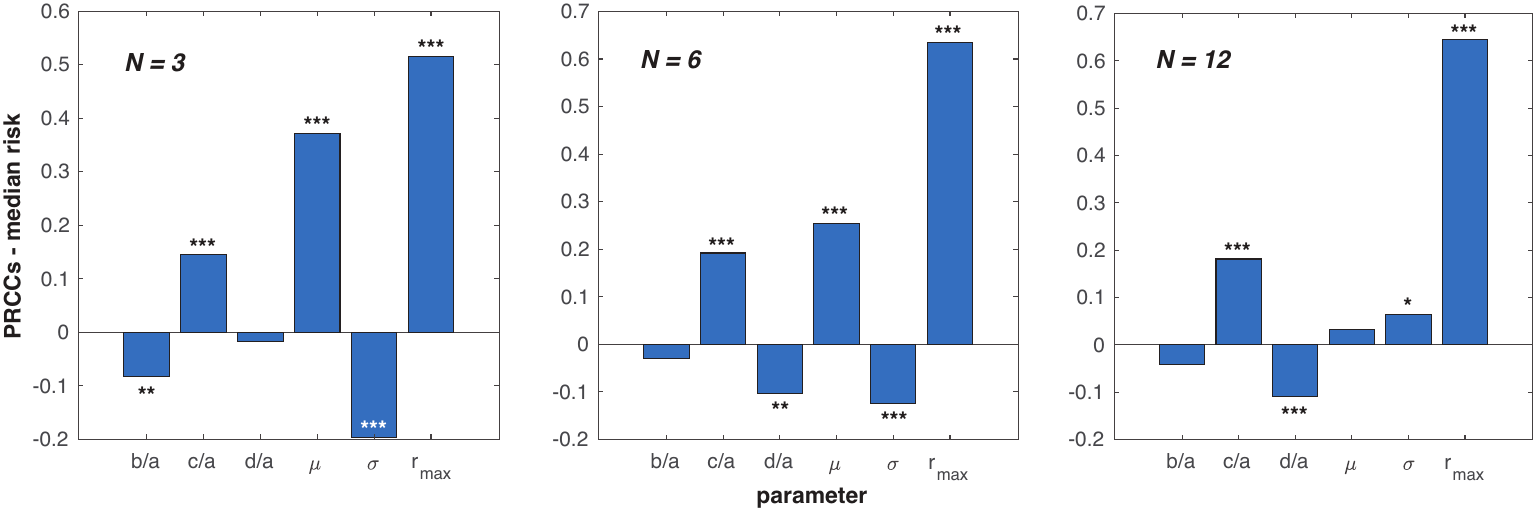}
      \caption{Sensitivity of parameters on the median steady state perceived risk level $r_i$. The sensitivity analysis uses Latin Hypercube Sampling (LHS) of parameter space and partial rank correlation coefficients (PRCC) \cite{marino2008methodology}. All parameter values are taken near the baselines in Table \ref{tab:param}. Initial conditions are sampled from $r_i(0) \sim \mathcal{U}(0,r_{\text{max}})$, where $r_{\text{max}}$ is also a parameter. Asterisks indicate significance with 1000 simulations ($^*p<0.05$, $^{**}p<0.01$, $^{***}p<0.001$). 
      We see that regardless of crowd size $N$, steady state risk levels are most sensitive to the maximum initial risk ($r_{\text{max}}$). For small crowd sizes, the distribution of loss aversion ratios impacts learned risk substantially; for large crowd sizes, learning rates have more influence on learned risk.} \label{fig:sensitivity}
\end{figure}

The sensitivity analysis (Figure \ref{fig:sensitivity}) indicates that, regardless of crowd size, steady state risk levels are most sensitive to the maximum initial risk ($r_{\text{max}}$), where the initial conditions are sampled from $r_i(0) \sim \mathcal{U}(0,r_{\text{max}})$. Larger median initial risk leads to larger median steady state risk. This occurs because individuals are less likely to intervene when society perceives an action as inappropriate; when people don't act, no one can learn from the outcomes of actions, and therefore risk remains high.

Also for any crowd size, increasing the learning rate from others' mistakes ($c$) positively impacts steady state risk; this is sensible because, if people learn more quickly from watching an intervenor embarrass themselves, then they will be less likely to intervene. The learning rates from positive bystander outcomes, when no one is ``called out'' have less impact on steady state risk; learning from others' positive outcomes ($d$) becomes more significant as crowd size grows (because more interventions occur from which to learn), but learning from one's own positive outcomes ($b$) is only significant for small crowd sizes (because one individual is less likely to act in a larger crowd). 

Interestingly, the distribution of loss aversion ratios most substantially impacts steady state risk for small crowd sizes but shrinks in both magnitude and significance as crowd size grows. For loss aversion ratios sampled from $\displaystyle x_i \sim \Gamma\left(\alpha = \left(\frac{\mu}{\sigma}\right)^2,\beta = \frac{\sigma^2}{\mu}\right)$, the mean of the distribution positively influences steady state risk and the standard deviation negatively influences risk, but these impacts shrink as the crowd size grows. Plausibly, this phenomenon occurs because really large crowds (here, a dozen or more) discourage action for most witnesses regardless of their loss aversion; learning from the few witnesses who risk intervention is far more important.

\subsection{Data} \label{sec:data}
To validate our model, we compile a database of observational and experimental data from studies of the bystander effect and how it manifests in a variety of conditions. Literature searches were conducted in databases PubMed, SageJournals, APA PsycNet, JSTOR, Wiley, and Google Scholar, utilizing keywords ``bystander intervention'', ``ambiguity'', ``group size'', ``diffusion of responsibility'', ``bystander apathy'', ``social influence'', ``model'', and various combinations of the terms. We include a study if it includes two critical components: (1) group size and (2) a quantitative indicator of intervention frequency\footnote{We exclude two studies in which the focal individual is a child too young to have developed theory of mind.}. 

We compute group size based on the number of victims, perpetrators, and witnesses present in the situation; witnesses were either focal individuals (i.e., subjects) or confederates (i.e., experimenters posing as subjects). In virtual situations, crowd sizes are not directly observed but instead indicated by numbers visible to the subjects. When ranges of crowd sizes are provided, we use the average value (for finite ranges) or the minimum value (for unbounded ranges). 

Intervention frequency is most often provided as either the fraction of situations in which the victim was assisted or the fraction of witnesses who intervened; when there is only one subject as the witness, these fractions are identical. A few studies provide intervention using another metric (e.g., average restaurant gratuity rate or average intention to intervene on a Likert scale). When reasonable, we convert these proxies into intervention fractions. 

\begin{figure}[t]
\includegraphics[width=0.49\textwidth]{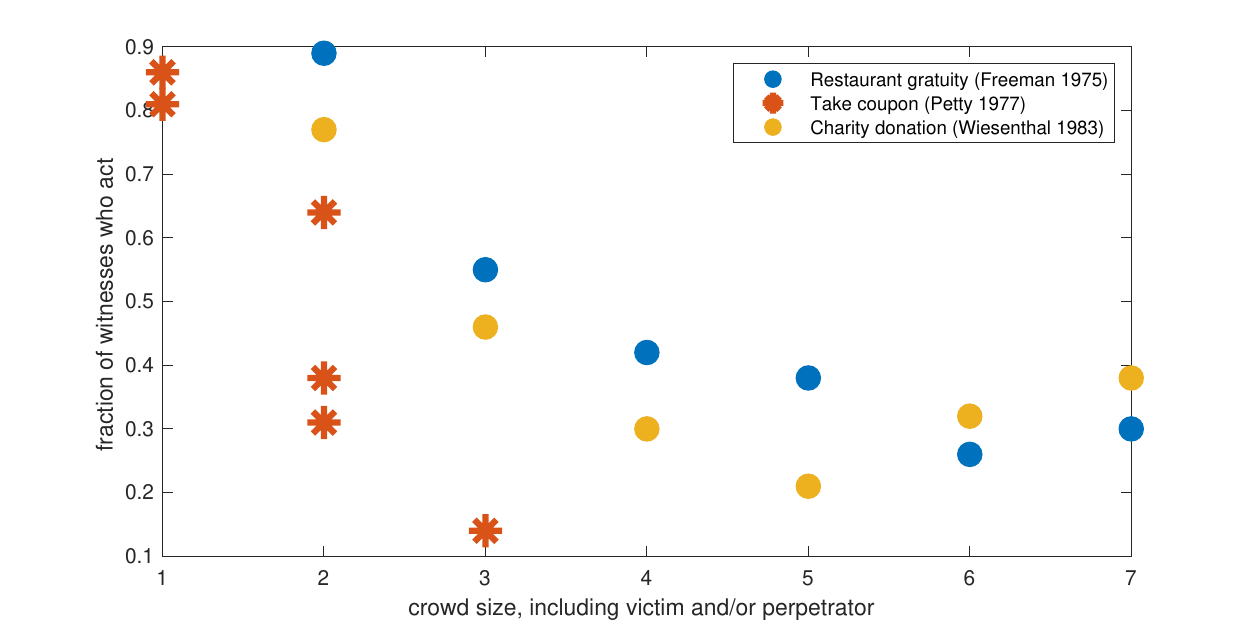}
\includegraphics[width=0.49\textwidth]{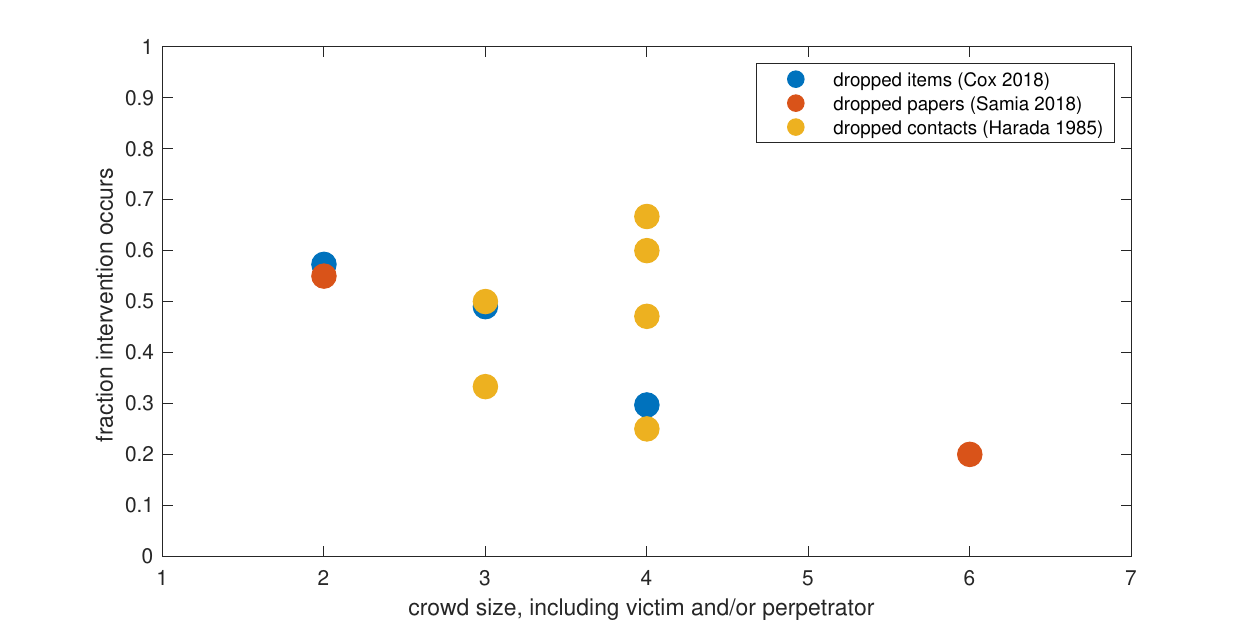} \\

\includegraphics[width=0.49\textwidth]{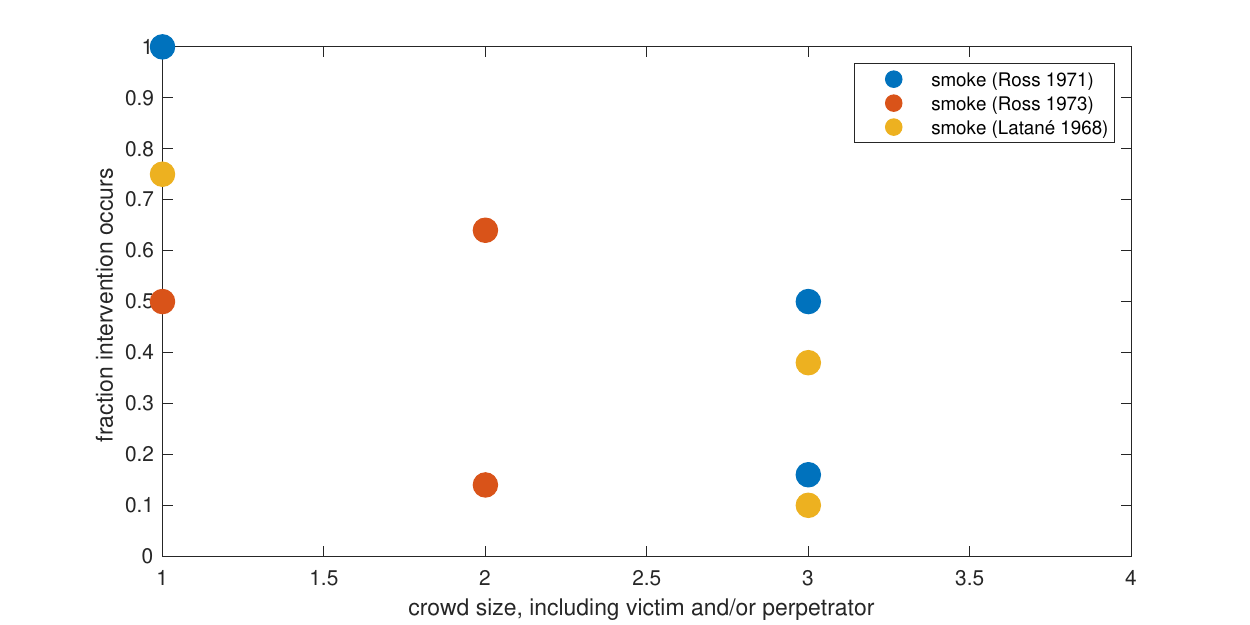}
\includegraphics[width=0.49\textwidth]{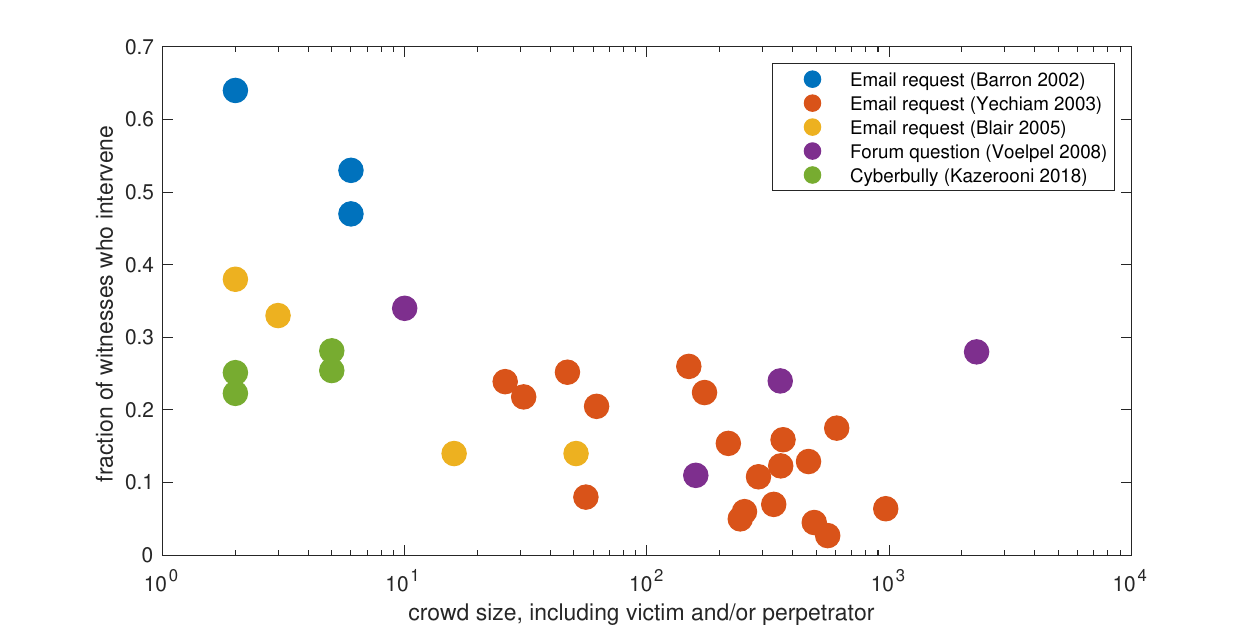}
\caption{Bystander situation data where a bystander effect may be visible, separated by the type of situation: financial situation (top left) \cite{freeman1975diffusion,petty1977social,wiesenthal1983diffusion}, dropped items (top right) \cite{cox2018bystander,samia2018number,harada1985ambiguity}, smoke in room (bottom left) \cite{ross1971effect,ross1973effect,latane1968inhibition}, virtual situation (bottom right) \cite{barron2002email,yechiam2003learning,blair2005electronic,voelpel2008david,kazerooni2018cyberbullying}. helping situation (top left). Qualitative bystander effects can plausibly be observed in non-dangerous situations, but a bystander effect is not clear in emergency situations. Due to the small number of data points for each study, we are unable to fit our model to each situation, although the trends are consistent with our model. Note that some studies provide intervention probability for individual witnesses, while other studies provide the probability that an intervention occurs (by any witness). The asterisk data points for the coupon study indicate that, unlike the rest of the studies in the plot, this study measured the probability that intervention occurred.}
\label{fig:data1}
\end{figure}

\begin{figure}[t]
\includegraphics[width=0.49\textwidth]{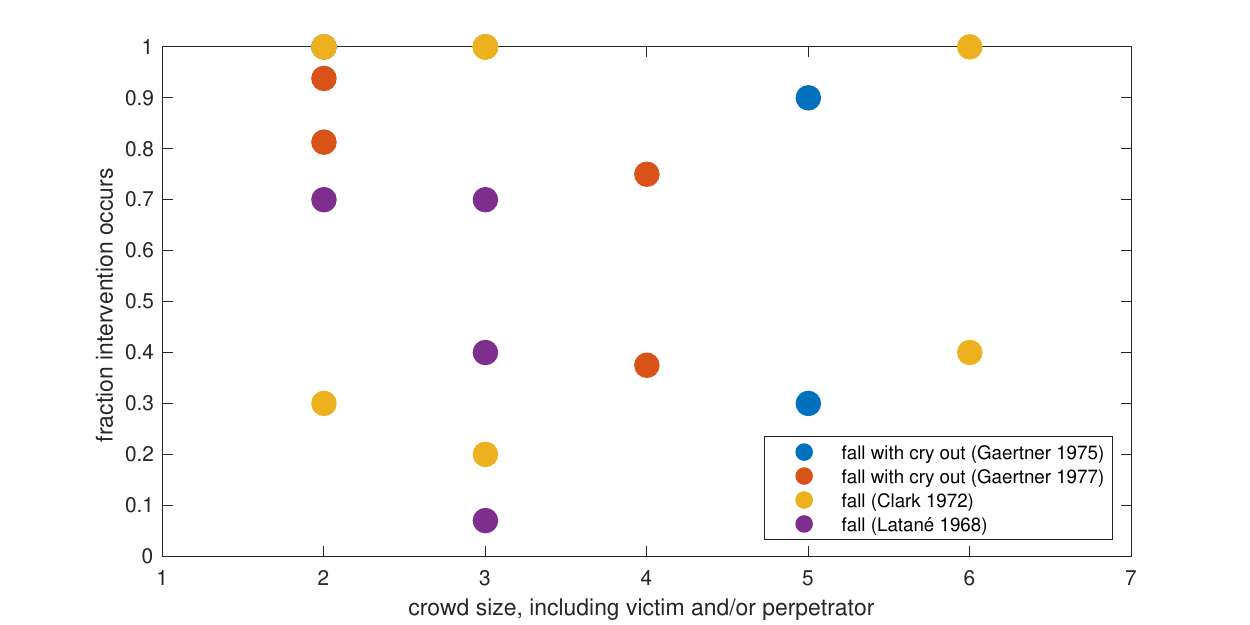}
\includegraphics[width=0.49\textwidth]{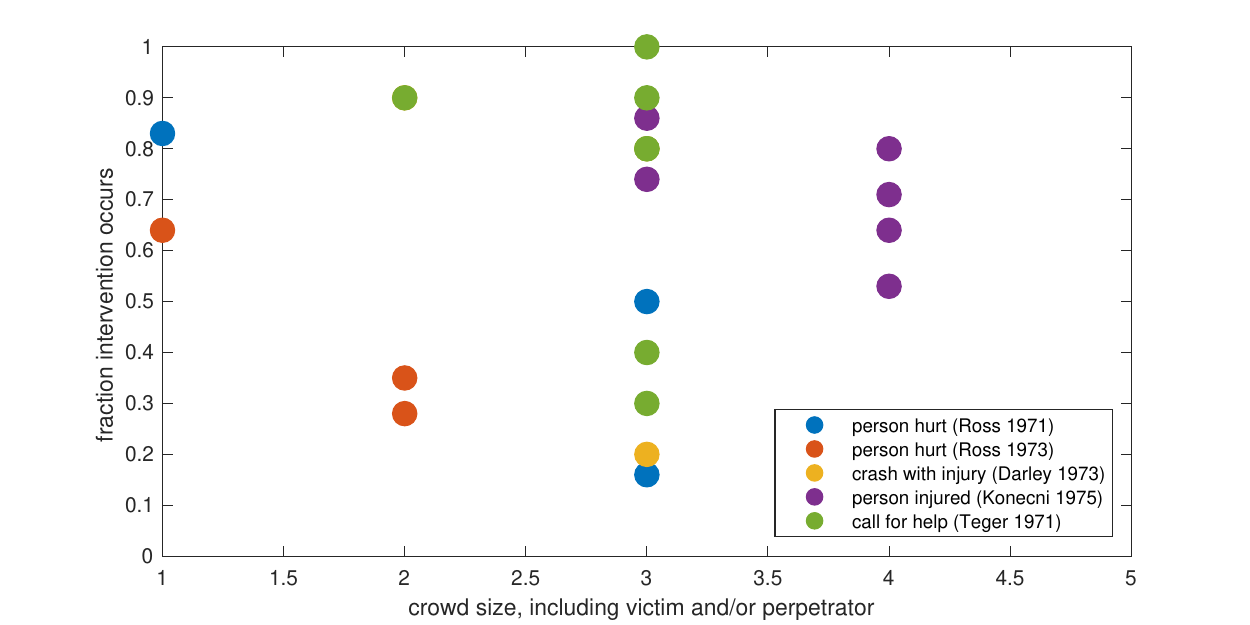} \\

\includegraphics[width=0.49\textwidth]{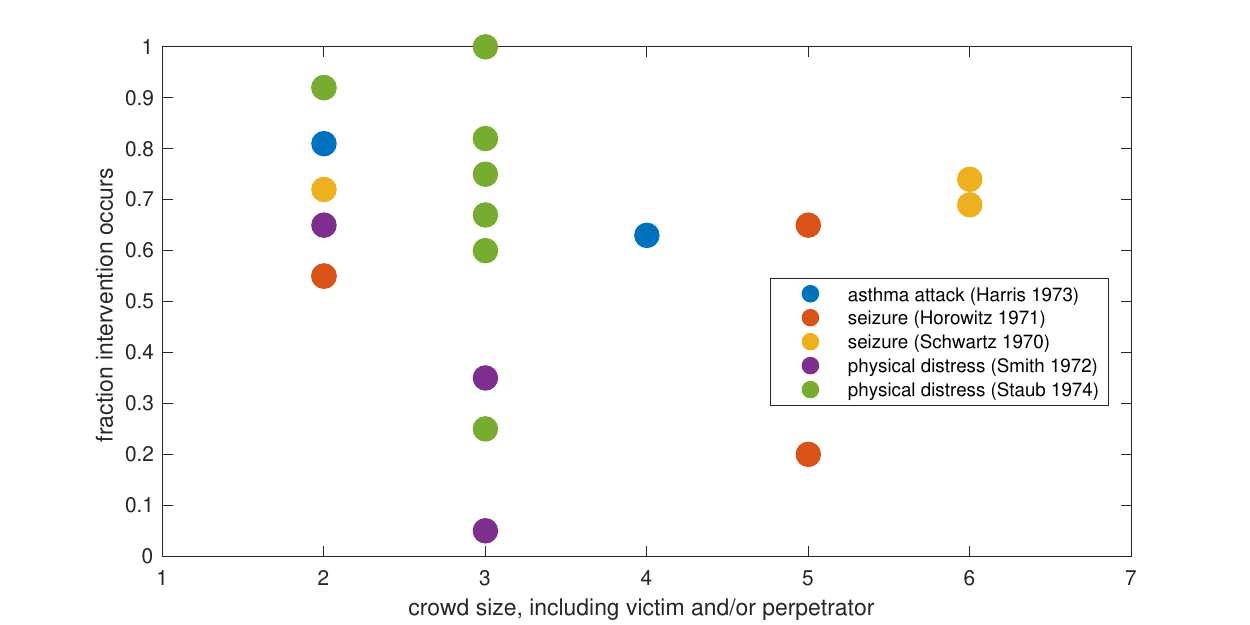}
\includegraphics[width=0.49\textwidth]{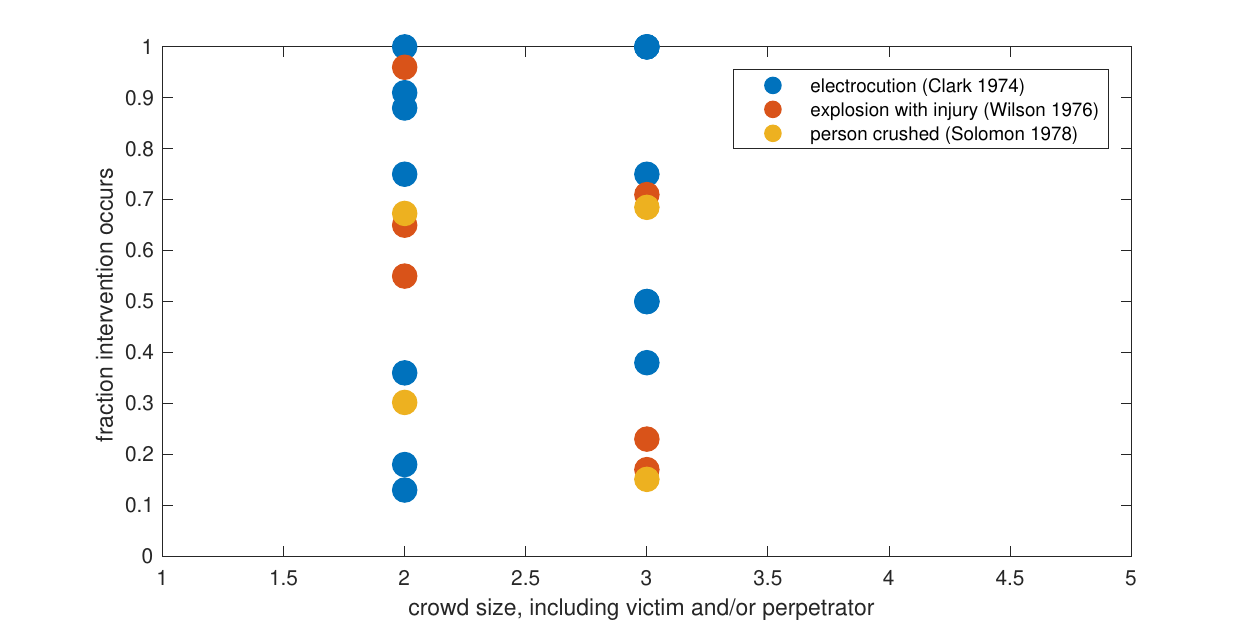} \\

\includegraphics[width=0.49\textwidth]{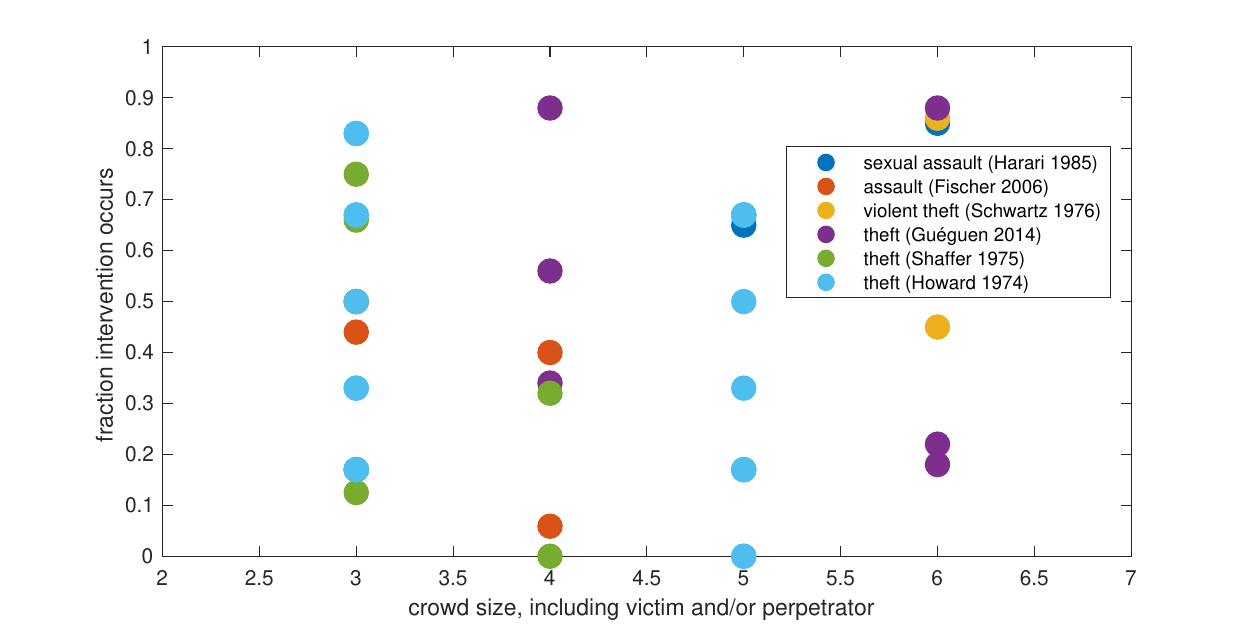}
\caption{Bystander situation data where a bystander effect may not be visible, separated by the type of situation: fall (top left) \cite{gaertner1975role,gaertner1977subtlety,clark1972don,latane1969lady}, mild injury (top right) \cite{ross1971effect,ross1973effect,darley1973groups,konevcni1975effects,teger1971examination}, medical emergency (middle left) \cite{harris1973bystander,horowitz1971effect,schwartz1970responsibility,smith1972inhibition,staub1974helping}, serious injury (middle right) \cite{clark1974apathetic,solomon1978helpingfunc,wilson1976motivation}, criminal situation (bottom) \cite{harari1985rape,fischer2006unresponsive,schwartz1976theft,gueguen2015commitment,shaffer1975intervention,howard1974effects}. Qualitative bystander effects can plausibly be observed in non-dangerous situations, but a bystander effect is not clear in emergency situations. Due to the small number of data points for each study, we are unable to fit our model to each situation, although the trends are consistent with our model. Note that some studies provide intervention probability for individual witnesses, while other studies provide the probability that an intervention occurs (by any witness). The asterisk data points for the coupon study indicate that, unlike the rest of the studies in the plot, this study measured the probability that intervention occurred.}
\label{fig:data2}
\end{figure}

We divide the studies into broad categories for the purpose of visualization: non-dangerous monetary situations (donating to charity, leaving a gratuity, etc.), dropped items, smoke seeping into room, virtual requests, falls either with a cry for help or not, injuries ranging from mild to serious, medical emergencies, and criminal offenses (theft, assault, etc.). See Figure \ref{fig:data1} for situations that appear to show a bystander effect and Figure \ref{fig:data2} for situations that don't appear to show a bystander effect.

\section{Results}
\begin{figure}[t]
\includegraphics[width=0.8\textwidth]{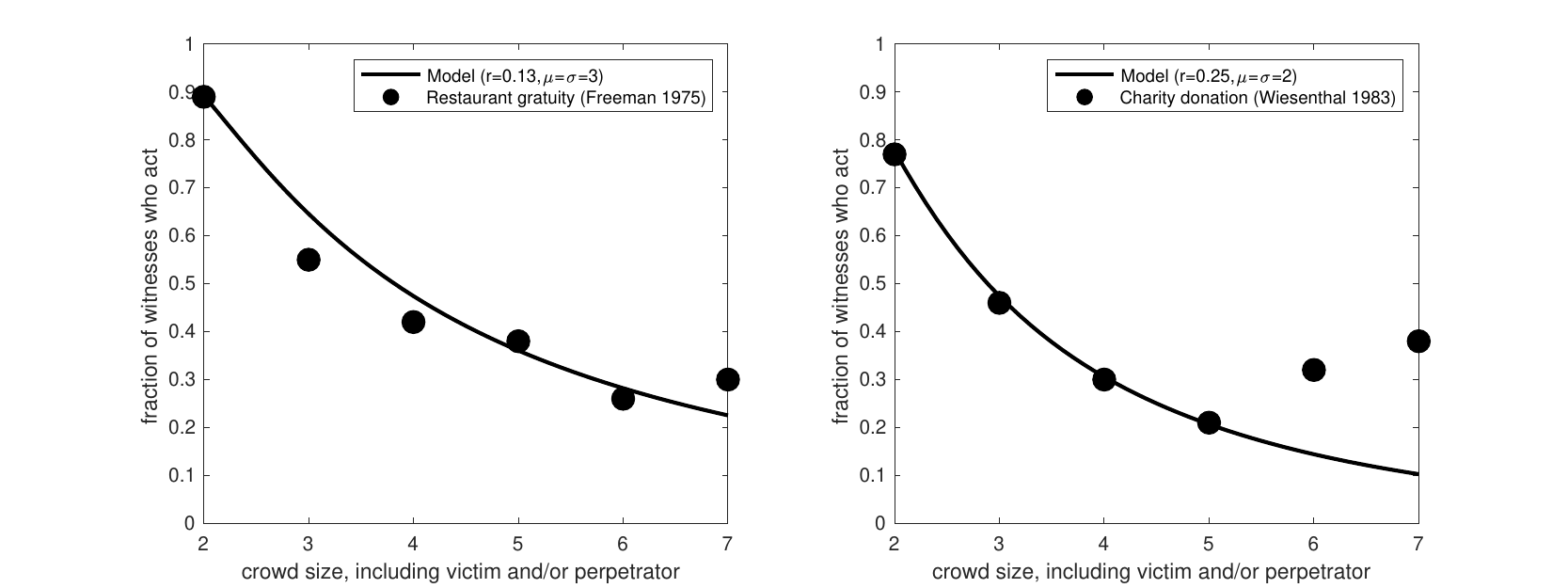} \\
\includegraphics[width=0.8\textwidth]{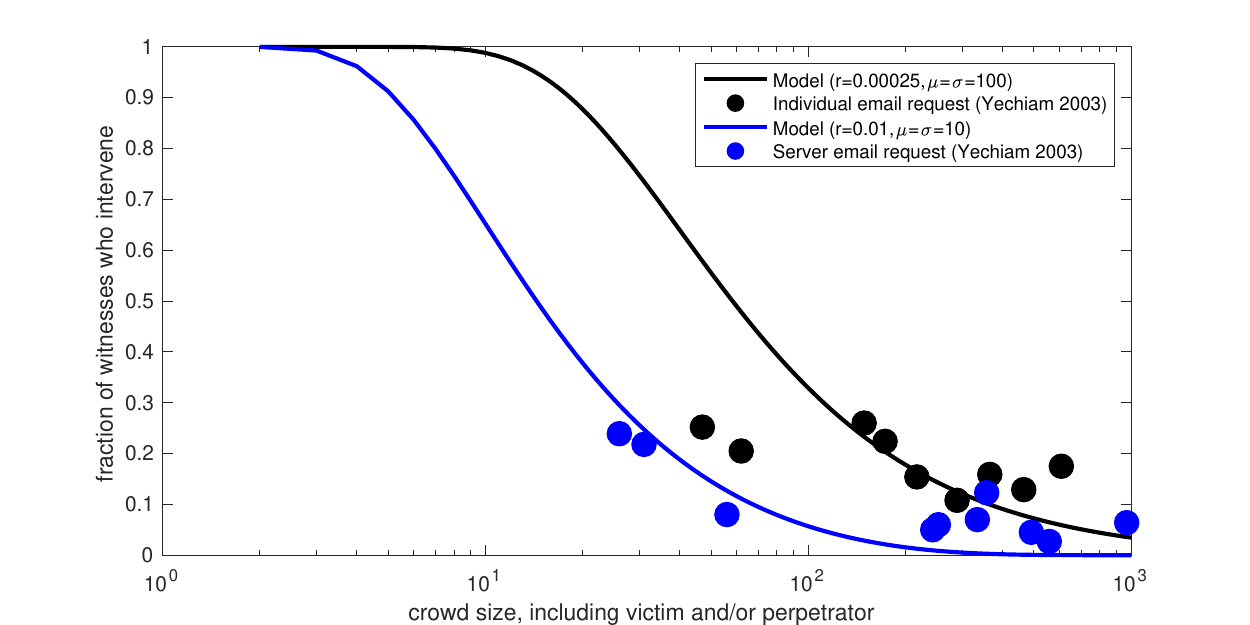}
\caption{Model plausibility using data from three different observational or experimental bystander situations: restaurant tip amount as a function of party size \cite{freeman1975diffusion}, linearly interpolating 20\% as all witnesses acting and 10\% as no witnesses acting (top left), fraction who donated to a relief fund as a function of group size at a public event \cite{wiesenthal1983diffusion} (top right), and fraction who responded to an email request \cite{yechiam2003learning} (bottom). The static model (\ref{eq:static}) is not fit due to practical identifiability issues, so model curves merely show plausible outcomes.}
\label{fig:valid}
\end{figure}

Because most bystander studies were designed for hypothesis testing, the majority of studies include only two or three different crowd sizes. Therefore, we cannot in good faith fit a mathematical model to each situation. One might reasonably decide to fit our model to ensembles of similar situations, but it is unclear what situations would have similar enough action appropriateness and loss aversion distributions. Because of data limitations, we refrain from fitting our model and simply show that our model is consistent with some situations and less consistent with other situations for studies that include at least six distinct crowd sizes (see Figure \ref{fig:valid}).

\begin{figure}[t]
\includegraphics[width=0.8\textwidth]{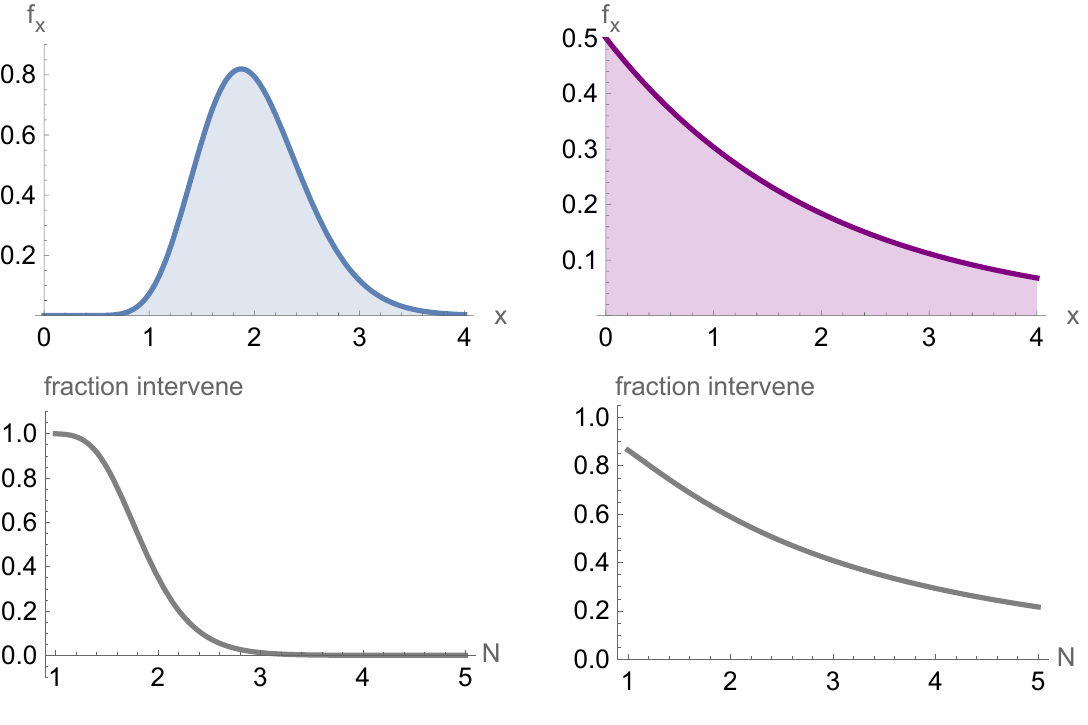}
\caption{Emergent bystander curves (lower) as a function of loss aversion distributions (upper). If the loss aversion ratio follows a gamma distribution with $\mu=2, \sigma=0.5$ (top left), then the bystander curve is sigmoidal (bottom left); this is not commonly seen in bystander studies. If the loss aversion ratio follows a gamma distribution with $\mu=2, \sigma=2$, i.e., an exponential distribution (top right), then the bystander curve is decaying (bottom right); this is more frequently seen in bystander studies.}
\label{fig:gammaexp}
\end{figure}

Although we cannot fit the model due to practical identifiability issues caused by small datasets, we can observe several general patterns. Among datasets that exhibit a bystander effect (e.g., non-dangerous situations), the fraction of witnesses who intervene tends to have a decreasing, concave up shape as a function of crowd size. In our static model (\ref{eq:static}) with assumed uniform action appropriateness, this occurs either when the risk of action is small but not negligible, or when variation in loss aversion is large (standard deviation approaches the mean of the assumed gamma distribution, leading to an exponential distribution). In the studies with at least six different crowd sizes, all informal fits required relatively low risk and large loss aversion variation (equal to the mean loss aversion) in order to be consistent with the data. This suggests that emotional prospects may have a larger variance than monetary loss aversion in the population (see Figure \ref{fig:gammaexp}). While model (\ref{eq:static}) fits the financial situations quite well, the model does not fit the virtual situations well; potentially, this occurs because the virtual situations use large numbers of virtual witnesses, which people may not be capable of comprehending \cite{trick1994small,lytle2021bystander}.

The data are also consistent with smaller variation in loss aversion if action appropriateness is not assumed uniform. Taking loss aversion to follow a gamma distribution with $\mu=2, \sigma=0.5$ \cite{arora2015risk, blake2021quantifying,kahneman2011thinking}, a uniform distribution of action risk $U(0,0.1)$ produces a similar decreasing, concave up shape seen in many of the datasets (see Figure \ref{fig:behav4}); this shape is substantially different from assuming all individuals hold the average risk level of the distribution. The data are less consistent with the steady state of the dynamic model (\ref{eq:odes}), which exaggerates the switch between a large and small intervention probability as crowd size grows (see Figure \ref{fig:behav4}). This suggests that social learning is slow enough to not reach steady state.

\section{Discussion}
Several qualitative mechanisms have been proposed for the observed bystander effect, most prominently diffusion of responsibility, audience inhibition, and social influence. The static model (\ref{eq:static}) presented here quantitatively tests the implications of the audience inhibition hypothesis, and the dynamic model (\ref{eq:odes}) incorporates social influence (although on a longer timescale than typically considered in the psychological literature). Either under a static or dynamic model, the individual fear of public embarrassment causes the collective intervention probability to decrease as crowd size grows. 

There is strong evidence that social embarrassment inhibits bystanders from acting, even when the situation is clear but the appropriateness of action may be ambiguous (e.g., a colleague has food in their teeth or ink on their face) \cite{zoccola2011embarrassed}. In fact, many studies have shown that inaction is preferred to action in many situations due to regret-induced status quo bias \cite{nicolle2011regret,han2023decision}; the fundamental lack of symmetry between action and inaction (i.e., status quo) is not captured by a simple model of diffusion of responsibility, but \textit{is} captured by our model of audience inhibition using loss aversion (prospect theory assumes the utility of inaction is zero, while the utility of a loss or gain due to an action is governed by a person's loss aversion ratio).

That said, many of the bystander datasets appear consistent with a simple model of diffusion of responsibility as well, e.g., the fraction who intervene is proportional to $1/N$ (or some power of $1/N$). With such limited data (usually only 2-3 data points per bystander study), it would be unreasonable to use model selection techniques (e.g., AIC or cross-validation) to choose between a model of audience inhibition and a model of diffusion of responsibility. However, there is evidence that simple diffusion of responsibility is not a universal explanation for the bystander effect, as bystanders report after some situations that they did not expect others to intervene \cite{schwartz1976theft}. Also, bystander effects are seen in non-helping situations (e.g., people are free to take a coupon from a limited supply), where diffusion of responsibility does not apply.

Finally, many of the bystander datasets may also be consistent with the social influence hypothesis, where action is more likely if others act and inaction is more likely if others fail to act. Our dynamic model (\ref{eq:odes}) captures the influence of others' behavior on future action or inaction in similar situations, although we do not directly model the cascade of action/inaction within a single bystander situation. A more sophisticated version of our model would break our current assumption that bystander decisions are independent within a single bystander situation. 

Assuming our model has captured the important features of bystander situation decisions, the bystander intervention curve (crowd size versus intervention probability) shows that a bystander effect consistent with non-dangerous bystander situations emerges only under certain circumstances: (1) risk of intervention is relatively low on average, (2) risk of intervention has some variability in the population, and/or (3) loss aversion has high variability in the population. The first condition is sensible because most bystander experiments are designed with an intervention that is deemed more culturally acceptable than not. The second and third conditions imply that heterogeneity in the population is a key driver of the observed bystander effect. If everyone agreed on the appropriateness of action and shared their aversion to loss, then each bystander situation would have a hard crowd-size threshold for intervention, below which everyone would act and above which no one would act. None of the bystander experiments show this behavior. 

\subsection{Limitations and Future Steps}
Like all minimal models, both the static (\ref{eq:static}) and dynamic (\ref{eq:odes}) models of the bystander effect have simplified reality. In the static model, we assume that bystanders simultaneously witness a situation and independently decide whether or not to act, an assumption also made in the review paper \cite{latane1981tenyears}. This may be a good approximation in some situations where a quick decision needs to be made (e.g., to respond to a person waving, or to interrupt a laptop theft), but less applicable when bystanders have time to consider their decisions and watch others' behavior (e.g., to respond to an email request). While we have included social influence through our dynamic social learning model, an extension of the static model could incorporate intervention dependence and time-to-response within a single bystander situation. 

The static model also assumes that people accurately assess the social risk of intervention based on their personal risk. However, people tend to overestimate small probabilities, whether for rare good events (e.g., winning the lottery) or rare bad events (e.g., being in a plane crash) \cite{kahneman1979prospect}. If we incorporate this aspect of prospect theory, our results would change quantitatively but not qualitatively, assuming the relationship between estimated risk and actual risk is monotonic (which has been demonstrated experimentally \cite{khaw2021individual}). 

We have also assumed that people accurately count the number of witnesses in a bystander situation; while humans are capable of quickly and accurately counting exact quantities up to about 4 or 5 items, counting becomes effortful and slower for larger quantities \cite{trick1994small}. This does not limit our results in most bystander situations with small crowd sizes, but it may make our model less applicable in some situations where people use qualitative measures of quantity (e.g., email requests with many recipients \cite{yechiam2003learning}). However, our model results are nearly identical beyond crowd sizes of 10-12 (see Figure \ref{fig:behav3}), so any error in measuring large crowd sizes will likely be negligible.

Another simplification made by our model is that witnesses treat all other individuals equally in a bystander situation, ignoring both that victims/perpetrators may play a different role and that there may exist personal relationships among people. To our knowledge, no studies have explored how witness behavior changes if the roles of the participants changes (e.g., witness is instead a perpetrator), probably because the situation itself changes. On the other hand, relationships among individuals have been studied; in one study of a helping bystander situation, lower intervention rates are observed when the bystanders know each other in ambiguous situations, but intervention rates among acquaintances are higher in clear-cut situations \cite{harada1985ambiguity}. In other words, two friends inhibit each other's helping when the situation is unclear, but support each other's helping when the situation is clear. Because most bystander studies involve strangers, our model does not capture this detail. 

Next, we consider only the emotional cost/benefit of action, ignoring the tangible cost of action (e.g., money, time, physical harm). In most of the bystander studies we compiled, the tangible costs are minimal (e.g., respond to friendly gesture \cite{jones2021increased} or reply to an email \cite{blair2005electronic,barron2002email}). However, some involve actual money (e.g., donate to a relief fund \cite{wiesenthal1983diffusion} or tip a server \cite{freeman1975diffusion}) or may require substantial time (e.g., help pick up a stack of dropped papers \cite{cox2018bystander} or search for a lost contact lens \cite{harada1985ambiguity}). If these monetary or time costs are independent of crowd size, then we could incorporate the costs by shifting the bystander curve (crowd size versus intervention probability) down by a certain percentage. If the costs depend on crowd size $N$ (e.g., dividing a lengthy task among witnesses), the percentage reduction in helping probability could depend on $N$ as long as the relationship between $N$ and individual cost was known. While monetary or time costs are certainly important considerations, our results would only change qualitatively if the costs drop more substantially than risk of intervention grows as crowd size increases; in this case, the bystander effect may disappear or even reverse after a certain crowd size. This effect is seen in certain emergency or dangerous bystander situations (e.g., sexual assault \cite{harari1985rape} or verbal/physical abuse \cite{fischer2006unresponsive}), where a marginal increase in the bystander intervention fraction may drastically decrease the physical risk to each bystander. A more general version of our model could incorporate a tangible cost function, but we leave that step for future work.

Another simplifying assumption is that loss aversion is static and not correlated with social learning rates or initial action appropriateness. However, it is known that loss aversion \cite{blake2021quantifying,dawson2023gender,arora2015risk} and embarrassability \cite{kelly1997assessment,singelis1995culture,hofmann2010cultural} are correlated with culture, gender, race/ethnicity, income, and class (among other demographic features). A more sophisticated version of our model would explore the impact of correlations among model parameters, even allowing the parameters to evolve over time.

We also ignore situations where action appropriateness may directly depend on crowd size; for example, if an intern emails an entire company asking for help, replying to that email may be less appropriate (e.g., intern should not receive help when bothering too many people) or may be more appropriate (e.g., reply to the intern to correct the behavior). Because this $N$-dependent effect would be culture- and situation-specific, it may not be possible to objectively model.

Finally, our models assume that crowd size is constant for an individual, but realistically people experience and learn from bystander situations with many different crowd sizes. Because the steady-state learned risk is qualitatively different in an all-to-all network and in a (static) small-world network, we expect that social learning on a dynamic social network structure (i.e., where each node's degree changes over time) would also have different results. We leave this extension for future work.

\section{Conclusion}
A bystander effect is the observation that helping behavior in social situations decreases as crowd size grows. The effect is not universal, tending to emerge when the situation is not life-threatening. While diffusion of responsibility may be a mechanism for the bystander effect in some clear-cut helping situations, social influence and audience inhibition are arguably more plausible mechanisms when the situation is ambiguous, and the ``right'' thing to do is not clear. 

We build two models of a bystander situation: a static model (\ref{eq:static}) of action risk via prospect theory capturing audience inhibition, and a dynamic model of social learning capturing social influence (albeit on a longer time-scale than originally proposed). The static model exhibits an emergent bystander effect and the dynamic model exacerbates the effect, with the features (e.g., average helping, rapidness of helping decline) of the effect dependent on learned action appropriateness and the distribution of loss aversion in the population. Although data limitations prevent model selection among the three main hypotheses for the bystander effect, the data are consistent with our models in most non-dangerous situations. 

From these models, we learn that the bystander effect consistent with the data emerges only under certain circumstances: (1) loss aversion has high variability in the population, (2) risk of intervention is relatively low on average, and (3) social learning is slow enough that it does not reach steady state. If an emergent bystander effect is not desirable, this model also suggests several potential interventions to mitigate the effect by reducing the risk of action: increase the benefit of acting (e.g., praise helpers for acting), decrease the cost of acting inappropriately (e.g., empathetically correct inappropriate behavior without shaming), or directly reduce the risk of acting (e.g., teach the public about appropriate actions through signage or public service announcements).   

\section{Data Availability}
All data (.xlsx file) and software (Matlab .m and Mathematica .nb files) will be made publicly available via the Harvard Dataverse.

\section{Acknowledgements}
The authors thank Neer Bhardwaj, Nupoor Gandhi, AJ Ingram, Yiheng Li, Jonathan Morales, and Dayo Ogunmodede for contributions to early exploration of the model. Thanks are also due to the Illinois Geometry Lab (now Illinois Mathematics Lab) and Mathways Grant No. DMS-1449269 (SMC) for research support.

%

\begin{thebibliography}{100}

\bibitem{BystanderApathy}
S.~A. Nida, ``Bystander apathy,'' 08 2020.

\bibitem{latane1968whenpplhelp}
J.~Darley and B.~Latan{\'e}, ``When will people help in a crisis?,'' {\em
  Psychology Today}, vol.~2, 01 1968.

\bibitem{latane1968inhibition}
B.~Latan{\'e} and J.~M. Darley, ``Group inhibition of bystander intervention in
  emergencies,'' {\em Journal of Personality and Social Psychology}, vol.~10,
  no.~3, pp.~215--221, 1968.

\bibitem{darley1968diffusion}
J.~M. Darley and B.~Latan{\'e}, ``Bystander intervention in emergencies:
  Diffusion of responsibility,'' {\em Journal of Personality and Social
  Psychology}, vol.~8, no.~4, pp.~377--383, 1968.

\bibitem{latane1969apathy}
B.~Latan{\'e} and J.~M. Darley, ``Bystander "apathy",'' {\em American
  Scientist}, vol.~57, no.~2, pp.~244--268, 1969.

\bibitem{fischer2011metaanalysis}
P.~Fischer, J.~I. Krueger, T.~Greitemeyer, C.~Vogrincic, A.~Kastenm{\"u}ller,
  D.~Frey, M.~Heene, M.~Wicher, and M.~Kainbacher, ``The bystander-effect: A
  meta-analytic review on bystander intervention in dangerous and non-dangerous
  emergencies.,'' {\em Psychological bulletin}, vol.~137, no.~4, pp.~517--537,
  2011.

\bibitem{staub1970child}
E.~Staub, ``A child in distress: The influence of age and number of witnesses
  on children's attempts to help.,'' {\em Journal of Personality and Social
  Psychology}, vol.~14, no.~2, p.~130, 1970.

\bibitem{plotner2015young}
M.~Pl{\"o}tner, H.~Over, M.~Carpenter, and M.~Tomasello, ``Young children show
  the bystander effect in helping situations,'' {\em Psychological science},
  vol.~26, no.~4, pp.~499--506, 2015.

\bibitem{bauman2020experiences}
S.~Bauman, J.~Yoon, C.~Iurino, and L.~Hackett, ``Experiences of adolescent
  witnesses to peer victimization: The bystander effect,'' {\em Journal of
  school psychology}, vol.~80, pp.~1--14, 2020.

\bibitem{gaertner1982race}
S.~L. Gaertner, J.~F. Dovidio, and G.~Johnson, ``Race of victim, nonresponsive
  bystanders, and helping behavior,'' {\em The Journal of Social Psychology},
  vol.~117, no.~1, pp.~69--77, 1982.

\bibitem{york2016racial}
E.~York~Cornwell and A.~Currit, ``Racial and social disparities in bystander
  support during medical emergencies on us streets,'' {\em American Journal of
  Public Health}, vol.~106, no.~6, pp.~1049--1051, 2016.

\bibitem{garcia2022racial}
R.~A. Garcia, J.~A. Spertus, S.~Girotra, B.~K. Nallamothu, K.~F. Kennedy, B.~F.
  McNally, K.~Breathett, M.~Del~Rios, C.~Sasson, and P.~S. Chan, ``Racial and
  ethnic differences in bystander cpr for witnessed cardiac arrest,'' {\em New
  England Journal of Medicine}, vol.~387, no.~17, pp.~1569--1578, 2022.

\bibitem{latane1975sex}
B.~Latan{\'e} and J.~M. Dabbs~Jr, ``Sex, group size and helping in three
  cities,'' {\em Sociometry}, pp.~180--194, 1975.

\bibitem{cox2018bystander}
A.~Cox and A.~Adam, ``The bystander effect in non-emergency situations:
  influence of gender and group size,'' {\em Modern Psychological Studies},
  vol.~23, no.~2, p.~3, 2018.

\bibitem{tice1985masculinity}
D.~M. Tice and R.~F. Baumeister, ``Masculinity inhibits helping in emergencies:
  Personality does predict the bystander effect.,'' {\em Journal of Personality
  and Social Psychology}, vol.~49, no.~2, p.~420, 1985.

\bibitem{jenkins2017bullying}
L.~N. Jenkins and A.~B. Nickerson, ``Bullying participant roles and gender as
  predictors of bystander intervention,'' {\em Aggressive behavior}, vol.~43,
  no.~3, pp.~281--290, 2017.

\bibitem{merrens1973nonemergency}
M.~R. Merrens, ``Nonemergency helping behavior in various sized communities,''
  {\em The Journal of Social Psychology}, vol.~90, no.~2, pp.~327--328, 1973.

\bibitem{shaffer1975intervention}
D.~R. Shaffer, M.~Rogle, and C.~Hendrlck, ``Intervention in the library: The
  effect of increased responsibility on bystanders' willingness to prevent a
  theft,'' {\em Journal of Applied Social Psychology}, vol.~5, no.~4,
  pp.~303--319, 1975.

\bibitem{latane1969lady}
B.~Latan{\'e} and J.~Rodin, ``A lady in distress: Inhibiting effects of friends
  and strangers on bystander intervention,'' {\em Journal of Experimental
  Social Psychology}, vol.~5, no.~2, pp.~189--202, 1969.

\bibitem{seo2022helping}
S.~Seo, T.~H. Witte, D.~M. Casper, and S.~Owen, ``Helping friends and strangers
  in risky situations: Outcomes of bystander interventions for sexual assault
  and dating violence,'' {\em Journal of Aggression, Maltreatment \& Trauma},
  vol.~31, no.~4, pp.~540--561, 2022.

\bibitem{bennett2016friends}
S.~Bennett and V.~L. Banyard, ``Do friends really help friends? the effect of
  relational factors and perceived severity on bystander perception of sexual
  violence.,'' {\em Psychology of violence}, vol.~6, no.~1, p.~64, 2016.

\bibitem{latane1981ten}
B.~Latan{\'e} and S.~Nida, ``Ten years of research on group size and
  helping.,'' {\em Psychological bulletin}, vol.~89, no.~2, p.~308, 1981.

\bibitem{krueger2009rational}
J.~I. Krueger and A.~L. Massey, ``A rational reconstruction of misbehavior,''
  {\em Social Cognition}, vol.~27, no.~5, pp.~786--812, 2009.

\bibitem{you2019bystander}
L.~You and Y.-H. Lee, ``The bystander effect in cyberbullying on social network
  sites: Anonymity, group size, and intervention intentions,'' {\em Telematics
  and Informatics}, vol.~45, p.~101284, 2019.

\bibitem{voelpel2008david}
S.~C. Voelpel, R.~A. Eckhoff, and J.~F{\"o}rster, ``David against goliath?
  group size and bystander effects in virtual knowledge sharing,'' {\em Human
  Relations}, vol.~61, no.~2, pp.~271--295, 2008.

\bibitem{blair2005electronic}
C.~A. Blair, L.~Foster~Thompson, and K.~L. Wuensch, ``Electronic helping
  behavior: The virtual presence of others makes a difference,'' {\em Basic and
  Applied Social Psychology}, vol.~27, no.~2, pp.~171--178, 2005.

\bibitem{markey2000bystander}
P.~M. Markey, ``Bystander intervention in computer-mediated communication,''
  {\em Computers in Human Behavior}, vol.~16, no.~2, pp.~183--188, 2000.

\bibitem{schwartz1976theft}
S.~H. Schwartz and A.~Gottlieb, ``Bystander reactions to a violent theft: Crime
  in jerusalem,'' {\em Journal of Personality and Social Psychology}, vol.~34,
  no.~6, pp.~1188--1199, 1976.

\bibitem{barron2002email}
G.~Barron and E.~Yechiam, ``Private e-mail requests and the diffusion of
  responsibility,'' {\em Computers in Human Behavior}, vol.~18, no.~5,
  pp.~507--520, 2002.

\bibitem{harari1985rape}
H.~Harari, O.~Harari, and R.~V. White, ``The reaction to rape by american male
  bystanders,'' {\em The Journal of Social Psychology}, vol.~125, no.~5,
  pp.~653--658, 1985.

\bibitem{fischer2006unresponsive}
P.~Fischer, T.~Greitemeyer, F.~Pollozek, and D.~Frey, ``The unresponsive
  bystander: Are bystanders more responsive in dangerous emergencies?,'' {\em
  European Journal of Social Psychology}, vol.~36, pp.~267--278, 2006.

\bibitem{franzen2013volunteer}
A.~Franzen, ``The volunteer's dilemma: Theoretical models and empirical
  evidence,'' in {\em Resolving Social Dilemmas}, pp.~135--148, Psychology
  Press, 2013.

\bibitem{clark1974apathetic}
R.~D. Clark and L.~E. Word, ``Where is the apathetic bystander? situational
  characteristics of the emergency.,'' {\em Journal of Personality and Social
  Psychology}, vol.~29, no.~3, p.~279, 1974.

\bibitem{solomon1978helping}
L.~Z. Solomon, H.~Solomon, and R.~Stone, ``Helping as a function of number of
  bystanders and ambiguity of emergency,'' {\em Personality and Social
  Psychology Bulletin}, vol.~4, no.~2, pp.~318--321, 1978.

\bibitem{karakashian2006fear}
L.~M. Karakashian, M.~I. Walter, A.~N. Christopher, and T.~Lucas, ``Fear of
  negative evaluation affects helping behavior: the bystander effect
  revisited.,'' {\em North American Journal of Psychology}, vol.~8, no.~1,
  2006.

\bibitem{siligato2024freezing}
E.~Siligato, G.~Iuele, M.~Barbera, F.~Bruno, G.~Tordonato, A.~Mautone, and
  A.~Rizzo, ``Freezing effect and bystander effect: overlaps and differences,''
  {\em Psych}, vol.~6, no.~1, pp.~273--287, 2024.

\bibitem{latane1970unresponsive}
B.~Latan{\'e} and J.~Darley, ``The unresponsive bystander: Why doesn’t he
  help?; 1970,'' {\em New York, Appleton-Century-Croft}, 1970.

\bibitem{piliavin1982cooperation}
J.~Piliavin, J.~F. Dovidio, S.~L. Gaertner, and R.~D. Clark~III, ``Cooperation
  and helping behavior,'' {\em Academic Press, New York, NY}, pp.~279--304,
  1982.

\bibitem{pilavin1991arousalcostreward}
J.~Dovidio, J.~Piliavin, S.~Gaertner, D.~Schroeder, and R.~III, ``The arousal:
  Cost-reward model and the process of intervention: A review of the
  evidence.,'' {\em Prosocial Behavior}, vol.~12, 01 1991.

\bibitem{van2020help}
J.~van~den Heuvel and J.~Treur, ``To help or not to help: A network modelling
  approach to the bystander effect,'' in {\em Biologically Inspired Cognitive
  Architectures Meeting}, pp.~527--540, Springer, 2020.

\bibitem{isada2016economic}
Y.~Isada, N.~Igaki, and A.~Shibata, ``An economic model of bystanders’
  behaviour,'' {\em Journal of Applied Mathematics and Physics}, vol.~4, no.~1,
  pp.~33--38, 2016.

\bibitem{freeman1975diffusion}
S.~Freeman, M.~R. Walker, R.~Borden, and B.~Latan{\'e}, ``Diffusion of
  responsibility and restaurant tipping: Cheaper by the bunch,'' {\em
  Personality and Social Psychology Bulletin}, vol.~1, no.~4, pp.~584--587,
  1975.

\bibitem{banerjee1992simple}
A.~V. Banerjee, ``A simple model of herd behavior,'' {\em The quarterly journal
  of economics}, vol.~107, no.~3, pp.~797--817, 1992.

\bibitem{frank2010microeconomics}
R.~H. Frank and E.~Cartwright, {\em Microeconomics and behavior}, vol.~8.
\newblock McGraw-Hill New York, 2010.

\bibitem{kahneman1979prospect}
D.~Kahneman and A.~Tversky, ``Prospect theory: An analysis of decision under
  risk,'' {\em Econometrica}, vol.~47, no.~2, pp.~263--291, 1979.

\bibitem{gueguen2015commitment}
N.~Gu{\'e}guen, M.~Dupr{\'e}, P.~Georget, and C.~S{\'e}n{\'e}meaud,
  ``Commitment, crime, and the responsive bystander: Effect of the commitment
  form and conformism,'' {\em Psychology, Crime \& Law}, vol.~21, no.~1,
  pp.~1--8, 2015.

\bibitem{kazerooni2018cyberbullying}
F.~Kazerooni, S.~H. Taylor, N.~N. Bazarova, and J.~Whitlock, ``Cyberbullying
  bystander intervention: The number of offenders and retweeting predict
  likelihood of helping a cyberbullying victim,'' {\em Journal of
  Computer-Mediated Communication}, vol.~23, no.~3, pp.~146--162, 2018.

\bibitem{arora2015risk}
M.~Arora and S.~Kumari, ``Risk taking in financial decisions as a function of
  age, gender: mediating role of loss aversion and regret,'' {\em International
  Journal of Applied Psychology}, vol.~5, no.~4, pp.~83--89, 2015.

\bibitem{blake2021quantifying}
D.~Blake, E.~Cannon, and D.~Wright, ``Quantifying loss aversion: evidence from
  a uk population survey,'' {\em Journal of Risk and Uncertainty}, vol.~63,
  no.~1, pp.~27--57, 2021.

\bibitem{kahneman2011thinking}
D.~Kahneman, {\em Thinking, fast and slow}.
\newblock macmillan, 2011.

\bibitem{hortensius2018empathy}
R.~Hortensius and B.~De~Gelder, ``From empathy to apathy: The bystander effect
  revisited,'' {\em Current directions in psychological science}, vol.~27,
  no.~4, pp.~249--256, 2018.

\bibitem{yin2023differential}
C.~Yin, B.~Li, and T.~Gao, ``Differential effects of reward and punishment on
  reinforcement-based motor learning and generalization,'' {\em Journal of
  neurophysiology}, vol.~130, no.~5, pp.~1150--1161, 2023.

\bibitem{wachter2009differential}
T.~W{\"a}chter, O.~V. Lungu, T.~Liu, D.~T. Willingham, and J.~Ashe,
  ``Differential effect of reward and punishment on procedural learning,'' {\em
  Journal of Neuroscience}, vol.~29, no.~2, pp.~436--443, 2009.

\bibitem{sidowski1956influence}
J.~B. Sidowski, L.~B. Wyckoff, and L.~Tabory, ``The influence of reinforcement
  and punishment in a minimal social situation.,'' {\em The Journal of Abnormal
  and Social Psychology}, vol.~52, no.~1, p.~115, 1956.

\bibitem{gregory2007effects}
F.~Gregory~Ashby and J.~R.~B. O’Brien, ``The effects of positive versus
  negative feedback on information-integration category learning,'' {\em
  Perception \& psychophysics}, vol.~69, no.~6, pp.~865--878, 2007.

\bibitem{jones2021increased}
D.~L. Jones, J.~D. Nelson, and B.~Opitz, ``Increased anxiety is associated with
  better learning from negative feedback,'' {\em Psychology Learning \&
  Teaching}, vol.~20, no.~1, pp.~76--90, 2021.

\bibitem{moraes2025unravelling}
E.~M. B.~d. Moraes, D.~R. d.~S. Carvalho, J.~Sandars, T.~M. Ozahata, R.~Patel,
  D.~Cecilio-Fernandes, and T.~M. Santos, ``Unravelling the differences between
  observation and active participation in simulation-based education,'' {\em
  Medical Teacher}, vol.~47, no.~6, pp.~991--996, 2025.

\bibitem{riley2017active}
J.~Riley and K.~Ward, ``Active learning, cooperative active learning, and
  passive learning methods in an accounting information systems course,'' {\em
  Issues in Accounting Education}, vol.~32, no.~2, pp.~1--16, 2017.

\bibitem{blanie2018impact}
A.~Blani{\'e}, S.~Gorse, P.~Roulleau, S.~Figueiredo, and D.~Benhamou, ``Impact
  of learners’ role (active participant-observer or observer only) on
  learning outcomes during high-fidelity simulation sessions in anaesthesia: a
  single center, prospective and randomised study,'' {\em Anaesthesia Critical
  Care \& Pain Medicine}, vol.~37, no.~5, pp.~417--422, 2018.

\bibitem{peruch2004active}
P.~P{\'e}ruch and P.~N. Wilson, ``Active versus passive learning and testing in
  a complex outside built environment,'' {\em Cognitive Processing}, vol.~5,
  no.~4, pp.~218--227, 2004.

\bibitem{eppinger2011choose}
B.~Eppinger and J.~Kray, ``To choose or to avoid: age differences in learning
  from positive and negative feedback,'' {\em Journal of Cognitive
  Neuroscience}, vol.~23, no.~1, pp.~41--52, 2011.

\bibitem{van2008evaluating}
A.~C. Van~Duijvenvoorde, K.~Zanolie, S.~A. Rombouts, M.~E. Raijmakers, and
  E.~A. Crone, ``Evaluating the negative or valuing the positive? neural
  mechanisms supporting feedback-based learning across development,'' {\em
  Journal of Neuroscience}, vol.~28, no.~38, pp.~9495--9503, 2008.

\bibitem{petzold2010stress}
A.~Petzold, F.~Plessow, T.~Goschke, and C.~Kirschbaum, ``Stress reduces use of
  negative feedback in a feedback-based learning task.,'' {\em Behavioral
  neuroscience}, vol.~124, no.~2, p.~248, 2010.

\bibitem{reime2017learning}
M.~H. Reime, T.~Johnsgaard, F.~I. Kvam, M.~Aarflot, J.~M. Engeberg, M.~Breivik,
  and G.~Bratteb{\o}, ``Learning by viewing versus learning by doing: A
  comparative study of observer and participant experiences during an
  interprofessional simulation training,'' {\em Journal of interprofessional
  care}, vol.~31, no.~1, pp.~51--58, 2017.

\bibitem{haidet2004controlled}
P.~Haidet, R.~O. Morgan, K.~O'malley, B.~J. Moran, and B.~F. Richards, ``A
  controlled trial of active versus passive learning strategies in a large
  group setting,'' {\em Advances in health sciences education}, vol.~9, no.~1,
  pp.~15--27, 2004.

\bibitem{chalmers1974learning}
D.~K. Chalmers and M.~E. Rosenbaum, ``Learning by observing versus learning by
  doing.,'' {\em Journal of Educational Psychology}, vol.~66, no.~2, p.~216,
  1974.

\bibitem{miller1987empathic}
R.~S. Miller, ``Empathic embarrassment: Situational and personal determinants
  of reactions to the embarrassment of another.,'' {\em Journal of Personality
  and social Psychology}, vol.~53, no.~6, p.~1061, 1987.

\bibitem{hawk2011taking}
S.~T. Hawk, A.~H. Fischer, and G.~A. Van~Kleef, ``Taking your place or matching
  your face: two paths to empathic embarrassment.,'' {\em Emotion}, vol.~11,
  no.~3, p.~502, 2011.

\bibitem{bernardo2008piecewise}
M.~Bernardo, C.~Budd, A.~R. Champneys, and P.~Kowalczyk, {\em Piecewise-smooth
  dynamical systems: theory and applications}, vol.~163.
\newblock Springer Science \& Business Media, 2008.

\bibitem{trick1994small}
L.~M. Trick and Z.~W. Pylyshyn, ``Why are small and large numbers enumerated
  differently? a limited-capacity preattentive stage in vision.,'' {\em
  Psychological review}, vol.~101, no.~1, p.~80, 1994.

\bibitem{marino2008methodology}
S.~Marino, I.~B. Hogue, C.~J. Ray, and D.~E. Kirschner, ``A methodology for
  performing global uncertainty and sensitivity analysis in systems biology,''
  {\em Journal of theoretical biology}, vol.~254, no.~1, pp.~178--196, 2008.

\bibitem{petty1977social}
R.~E. Petty, K.~D. Williams, S.~G. Harkins, and B.~Latan{\'e}, ``Social
  inhibition of helping yourself: Bystander response to a cheeseburger,'' {\em
  Personality and Social Psychology Bulletin}, vol.~3, no.~4, pp.~575--578,
  1977.

\bibitem{wiesenthal1983diffusion}
D.~L. Wiesenthal, D.~Austrom, and I.~Silverman, ``Diffusion of responsibility
  in charitable donations,'' {\em Basic and Applied Social Psychology}, vol.~4,
  no.~1, pp.~17--27, 1983.

\bibitem{samia2018number}
S.~Samia and M.~Khanam, ``Number of bystander, time pressure and gender effects
  on altruistic behavior,'' {\em Dhaka University Journal of Biological
  Sciences}, vol.~27, no.~1, pp.~1--7, 2018.

\bibitem{harada1985ambiguity}
J.~Harada, ``Bystander intervention: The effect of ambiguity of the helping
  situation and the interpersonal relationship between bystanders,'' {\em
  Japanese Psychological Research}, vol.~27, no.~4, pp.~177--184, 1985.

\bibitem{ross1971effect}
A.~S. Ross, ``Effect of increased responsibility on bystander intervention: The
  presence of children.,'' {\em Journal of personality and social psychology},
  vol.~19, no.~3, p.~306, 1971.

\bibitem{ross1973effect}
A.~S. Ross and J.~Braband, ``Effect of increased responsibility on bystander
  intervention: Ii. the cue value of a blind person.,'' {\em Journal of
  Personality and Social Psychology}, vol.~25, no.~2, p.~254, 1973.

\bibitem{yechiam2003learning}
E.~Yechiam and G.~Barron, ``Learning to ignore online help requests,'' {\em
  Computational \& Mathematical Organization Theory}, vol.~9, no.~4,
  pp.~327--339, 2003.

\bibitem{gaertner1975role}
S.~L. Gaertner, ``The role of racial attitudes in helping behavior,'' {\em The
  Journal of Social Psychology}, vol.~97, no.~1, pp.~95--101, 1975.

\bibitem{gaertner1977subtlety}
S.~L. Gaertner and J.~F. Dovidio, ``The subtlety of white racism, arousal, and
  helping behavior.,'' {\em Journal of Personality and social Psychology},
  vol.~35, no.~10, p.~691, 1977.

\bibitem{clark1972don}
R.~D. Clark and L.~E. Word, ``Why don't bystanders help? because of
  ambiguity?,'' {\em Journal of personality and social psychology}, vol.~24,
  no.~3, p.~392, 1972.

\bibitem{darley1973groups}
J.~M. Darley, A.~I. Teger, and L.~D. Lewis, ``Do groups always inhibit
  individuals' responses to potential emergencies?,'' {\em Journal of
  Personality and Social Psychology}, vol.~26, no.~3, p.~395, 1973.

\bibitem{konevcni1975effects}
V.~J. Kone{\v{c}}ni and E.~B. Ebbesen, ``Effects of the presence of children on
  adults' helping behavior and compliance: Two field studies,'' {\em The
  Journal of Social Psychology}, vol.~97, no.~2, pp.~181--193, 1975.

\bibitem{teger1971examination}
A.~I. Teger {\em et~al.}, ``An examination of the social influence hypothesis
  of bystander intervention in emergencies.,'' 1971.

\bibitem{harris1973bystander}
V.~A. Harris and C.~E. Robinson, ``Bystander intervention: Group size and
  victim status,'' {\em Bulletin of the Psychonomic Society}, vol.~2, no.~1,
  pp.~8--10, 1973.

\bibitem{horowitz1971effect}
I.~A. Horowitz, ``The effect of group norms on bystander intervention,'' {\em
  The Journal of Social Psychology}, vol.~83, no.~2, pp.~265--273, 1971.

\bibitem{schwartz1970responsibility}
S.~H. Schwartz and G.~T. Clausen, ``Responsibility, norms, and helping in an
  emergency.,'' {\em Journal of personality and social psychology}, vol.~16,
  no.~2, p.~299, 1970.

\bibitem{smith1972inhibition}
R.~E. Smith, L.~Smythe, and D.~Lien, ``Inhibition of helping behavior by a
  similar or dissimilar nonreactive fellow bystander.,'' 1972.

\bibitem{staub1974helping}
E.~Staub, ``Helping a distressed person: Social, personality, and stimulus
  determinants,'' in {\em Advances in experimental social psychology}, vol.~7,
  pp.~293--341, Elsevier, 1974.

\bibitem{solomon1978helpingfunc}
L.~Z. Solomon, H.~Solomon, and R.~Stone, ``Helping as a function of number of
  bystanders and ambiguity of emergency,'' {\em Personality and Social
  Psychology Bulletin}, vol.~4, no.~2, pp.~318--321, 1978.

\bibitem{wilson1976motivation}
J.~P. Wilson, ``Motivation, modeling, and altruism: A person$\times$ situation
  analysis.,'' {\em Journal of Personality and Social Psychology}, vol.~34,
  no.~6, p.~1078, 1976.

\bibitem{howard1974effects}
W.~Howard and W.~D. Crano, ``Effects of sex, conversation, location, and size
  of observer group on bystander intervention in a high risk situation,'' {\em
  Sociometry}, pp.~491--507, 1974.

\bibitem{lytle2021bystander}
R.~D. Lytle, T.~M. Bratton, and H.~K. Hudson, ``Bystander apathy and
  intervention in the era of social media,'' in {\em The emerald international
  handbook of technology-facilitated violence and abuse}, pp.~711--728, Emerald
  Publishing Limited, 2021.

\bibitem{zoccola2011embarrassed}
P.~M. Zoccola, M.~C. Green, E.~Karoutsos, S.~M. Katona, and J.~Sabini, ``The
  embarrassed bystander: Embarrassability and the inhibition of helping,'' {\em
  Personality and Individual Differences}, vol.~51, no.~8, pp.~925--929, 2011.

\bibitem{nicolle2011regret}
A.~Nicolle, S.~M. Fleming, D.~R. Bach, J.~Driver, and R.~J. Dolan, ``A
  regret-induced status quo bias,'' {\em Journal of Neuroscience}, vol.~31,
  no.~9, pp.~3320--3327, 2011.

\bibitem{han2023decision}
Q.~Han, S.~Quadflieg, and C.~J. Ludwig, ``Decision avoidance and post-decision
  regret: A systematic review and meta-analysis,'' {\em Plos one}, vol.~18,
  no.~10, p.~e0292857, 2023.

\bibitem{latane1981tenyears}
B.~Latan{\'e} and S.~Nida, ``Ten years of research on group size and helping,''
  {\em Psychological Bulletin}, vol.~89, no.~2, pp.~308--324, 1981.

\bibitem{khaw2021individual}
M.~W. Khaw, L.~Stevens, and M.~Woodford, ``Individual differences in the
  perception of probability,'' {\em PLoS computational biology}, vol.~17,
  no.~4, p.~e1008871, 2021.

\bibitem{dawson2023gender}
C.~Dawson, ``Gender differences in optimism, loss aversion and attitudes
  towards risk,'' {\em British Journal of Psychology}, vol.~114, no.~4,
  pp.~928--944, 2023.

\bibitem{kelly1997assessment}
K.~M. Kelly and W.~H. Jones, ``Assessment of dispositional embarrassability,''
  {\em Anxiety, Stress, and Coping}, vol.~10, no.~4, pp.~307--333, 1997.

\bibitem{singelis1995culture}
T.~M. Singelis and W.~F. Sharkey, ``Culture, self-construal, and
  embarrassability,'' {\em Journal of cross-cultural psychology}, vol.~26,
  no.~6, pp.~622--644, 1995.

\bibitem{hofmann2010cultural}
S.~G. Hofmann, M.~Anu~Asnaani, and D.~E. Hinton, ``Cultural aspects in social
  anxiety and social anxiety disorder,'' {\em Depression and anxiety}, vol.~27,
  no.~12, pp.~1117--1127, 2010.

\bibitem{watts1998collective}
D.~J. Watts and S.~H. Strogatz, ``Collective dynamics of
  ‘small-world’networks,'' {\em nature}, vol.~393, no.~6684, pp.~440--442,
  1998.

\bibitem{marinoCode}
D.~Kirschner, ``Uncertainty and sensitivity functions and implementation,''
  2008.

\bibitem{mckay1979comparison}
M.~McKay, R.~Beckman, and W.~Conover, ``Comparison of three methods for
  selecting values of input variables in the analysis of output from a computer
  cod,'' {\em Technometrics}, vol.~21, no.~2, p.~239–245, 1979.

\bibitem{saltelli2002making}
A.~Saltelli, ``Making best use of model evaluations to compute sensitivity
  indices,'' {\em Computer physics communications}, vol.~145, no.~2,
  pp.~280--297, 2002.

\end{thebibliography}

\setcounter{section}{0}
\renewcommand{\thesection}{S\arabic{section}}
\setcounter{equation}{0}
\renewcommand{\theequation}{S\arabic{equation}}
\setcounter{figure}{0}
\renewcommand{\thefigure}{S\arabic{figure}}
\setcounter{table}{0}
\renewcommand{\thetable}{S\arabic{table}}

\newpage

\section{Supplementary Materials}
\subsection{Probability Derivation}
\begin{figure}[th]
  \centering
    \includegraphics[width=\textwidth]{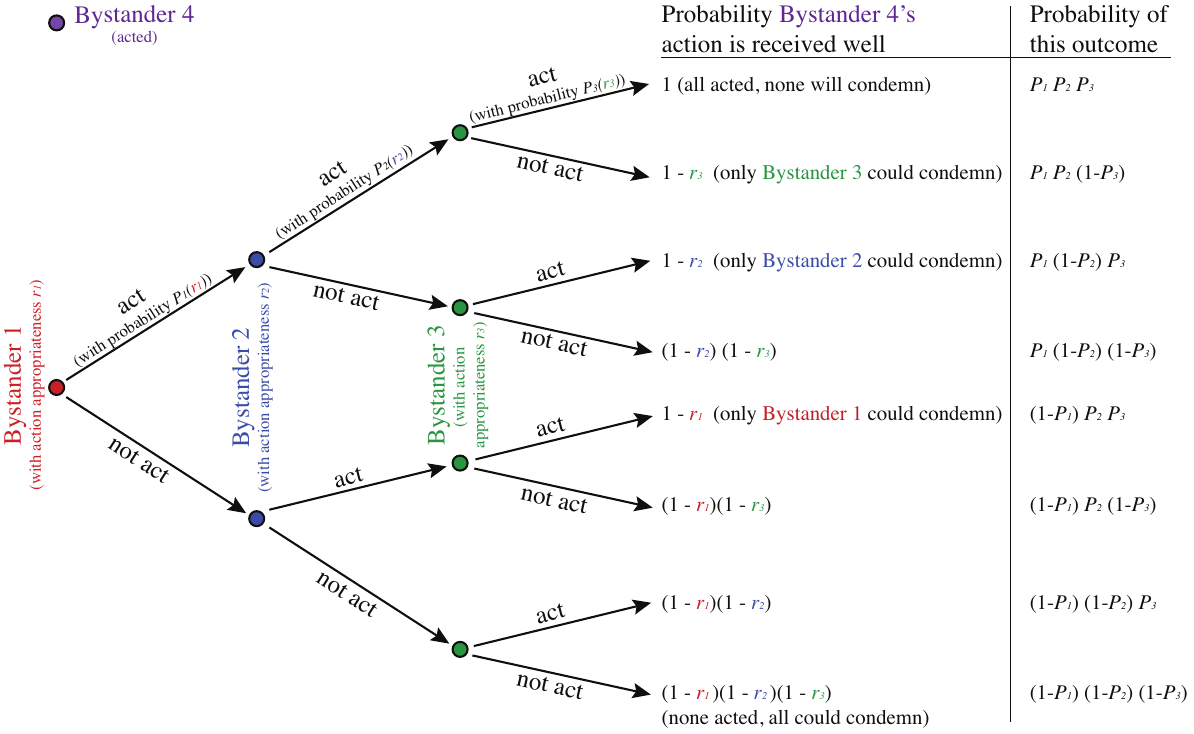}
      \caption{Example probability tree diagram with four bystanders, one of whom has decided to act. All probabilities of acting $P_i$ are assumed to be independent, and they depend on each action risk ($r_i$).} 
      \label{fig:tree-diagram}
\end{figure}

We state that the probability that individual $i$'s action is well-received, given that the individual acted, is
\begin{align*}
    P_i^{(w)} = \prod_{j\ne i} \bigg( P_j^{(I)} + \big(1-P_j^{(I)}\big) (1-r_j) \bigg)
\end{align*}
We illustrate the derivation of this probability using an example with four bystanders, focusing on a bystander (call them ``Bystander 4'') who has already decided to intervene. From the certainty that Bystander 4 has acted, we draw all possible outcomes for the other three bystanders (Figure \ref{fig:tree-diagram}). Note that we will suppress the superscript $(I)$ notation for all intervention probabilities to avoid notational clutter.

The probability that the intervention by Bystander 4 is received well in each possible branch is the product of all the risks held by the bystanders who did not act; if all bystanders acted (top branch), then no one will condemn an action and the probability action is well-received is 1. The probability that any particular branch is the eventual outcome is the product of all the probabilities of action or inaction. The overall probability of an action being well-received is therefore the sum of all the conditional probabilities:
\begin{align*}
P_4^{(w)} = \, & 1 \cdot P_1 P_2 P_3 + (1-r_3) \cdot P_1 P_2 (1-P_3) \\
            \, & + (1-r_2) \cdot P_1 (1-P_2) P_3 + (1-r_2) (1-r_3) \cdot P_1 (1-P_2) (1-P_3) \\
           \,  & + (1-r_1) \cdot (1-P_1) P_2 P_3 + (1-r_1) (1-r_3) \cdot (1-P_1) P_2 (1-P_3) \\
           \,  & + (1-r_1) (1-r_2) \cdot (1-P_1) (1-P_2) P_3 + (1-r_1) (1-r_2) (1-r_3) \cdot (1-P_1)(1-P_2)(1-P_3)
\end{align*}
We rearrange by factoring out $P_1$ and $(1-P_1)$ wherever possible and gathering like terms:
\begin{align*}
P_4^{(w)} = \, & P_1 \bigg[P_2 P_3 + (1-r_3) \cdot P_2 (1-P_3) + (1-r_2) \cdot (1-P_2) P_3 + (1-r_2) (1-r_3) \cdot (1-P_2) (1-P_3)\bigg] \\
           \,  & + (1-P_1) \bigg[(1-r_1) \cdot P_2 P_3 + (1-r_1) (1-r_3) \cdot P_2 (1-P_3) \\
           \, & \hspace{2cm} +(1-r_1) (1-r_2) \cdot (1-P_2) P_3 + (1-r_1) (1-r_2) (1-r_3) \cdot (1-P_2)(1-P_3)\bigg]
\end{align*}
Recursively, we continue this process (factoring out terms involving $P_2$ and then terms involving $P_3$, until we get to our last witness probability), yielding the following structure:
\begin{align*}
P_4^{(w)} = \, & P_1 \bigg[P_2 \Big[P_3 + (1-r_3) \cdot (1-P_3)\Big] + (1-P_2) \Big[(1-r_2) \cdot P_3 + (1-r_2) (1-r_3) \cdot(1-P_3)\Big]\bigg] \\
           \,  & + (1-P_1) \bigg[ P_2 \Big[(1-r_1) \cdot P_3 + (1-r_1) (1-r_3) \cdot (1-P_3)\Big] \\
           \, &  \hspace{2cm} + (1-P_2) \Big[ (1-r_1) (1-r_2) \cdot P_3 + (1-r_1) (1-r_2) (1-r_3) \cdot (1-P_3)\Big]\bigg]
\end{align*}
Then, we factor out all risk terms $(1-r_i)$, where possible:
\begin{align*}
P_4^{(w)} = \, & P_1 \bigg[P_2 \Big[P_3 + (1-r_3) (1-P_3)\Big] + (1-r_2)(1-P_2) \Big[P_3 + (1-r_3)(1-P_3)\Big]\bigg] \\
           + \,  & (1-r_1)(1-P_1) \bigg[ P_2 \Big[P_3 + (1-r_3) (1-P_3)\Big] + (1-r_2)(1-P_2) \Big[P_3 + (1-r_3) (1-P_3)\Big]\bigg]
\end{align*}
Finally, recursively factor out the common $\ds P_3 + (1-r_3) (1-P_3)$ term, and then the common $\ds P_2 + (1-r_2) (1-P_2)$ term: 
\begin{align*}
P_4^{(w)} = \, & \Big[ P_1 + (1-r_1) (1-P_1) \Big] \Big[ P_2 + (1-r_2) (1-P_2) \Big] \Big[ P_3 + (1-r_3) (1-P_3) \Big],
\end{align*}
which generalizes to our derived formula. 

We also claimed that the probability that another witness's action is well-received, given that individual $i$ did not intervene, is
\begin{align*}
    \tilde{P}_i^{(w)} = \prod_{j\ne i} \bigg( P_j^{(I)} + \big(1-P_j^{(I)}\big) (1-r_j) \bigg) - \prod_{j\ne i} \big(1-P_j^{(I)}\big) (1-r_j)
\end{align*}
The first product follows the same derivation as above. The second product corrects for the outcome in which no one acts, and therefore individual $i$ cannot learn from the situation. This outcome is not considered in the first derivation because individual $i$ was assumed to have acted.

\subsection{Social Learning on a Small World Network} \label{sec:model2}
\begin{figure}[th]
  \centering
    \includegraphics[width=\textwidth]{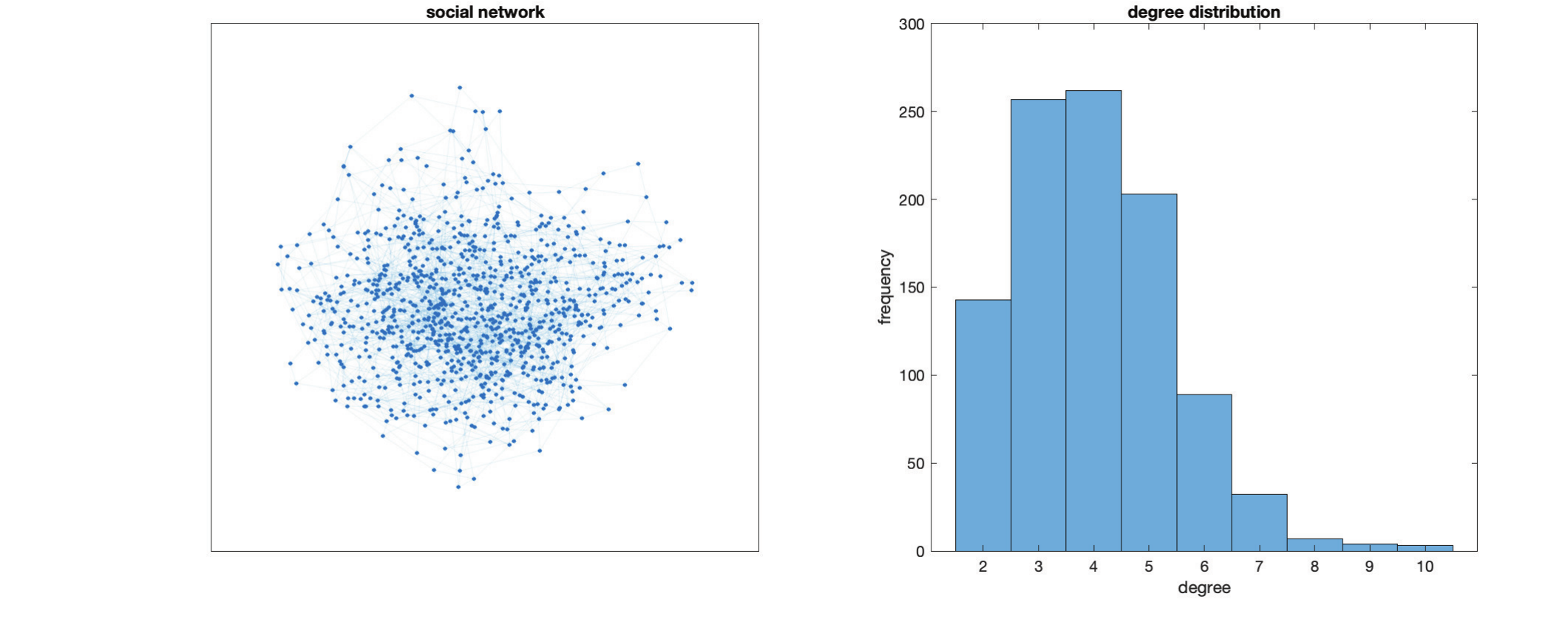}
      \caption{Small world network with 1000 nodes initially connected to $k=4$ neighbors with rewiring probability $\beta=1$ (left). After rewiring, the degree distribution (right) spans around 2 to 10 ($N$ spans 3 to 11).} \label{fig:network}
\end{figure}

To test the robustness of our all-to-all network results in Section \ref{sec:dynamicbehavior}, we investigate social learning on a small world network\footnote{Results were qualitatively similar with a scale free network.} \cite{watts1998collective} with $n = 1000$ individuals initially connected to $k=4$ neighbors with rewiring probability $\beta=1$, in order to achieve a degree distribution that (typically) spans $N=2$ or $3$ to $N=11$ or $12$. Each individual (node) on the network follows the dynamics of Equation \eqref{eq:odes}, learning only from their neighbors; this leads to a distribution of crowd sizes $N$ (degree$+1$) within the population (see Figure \ref{fig:network}).

\begin{figure}[th]
  \centering
    \includegraphics[width=0.85\textwidth]{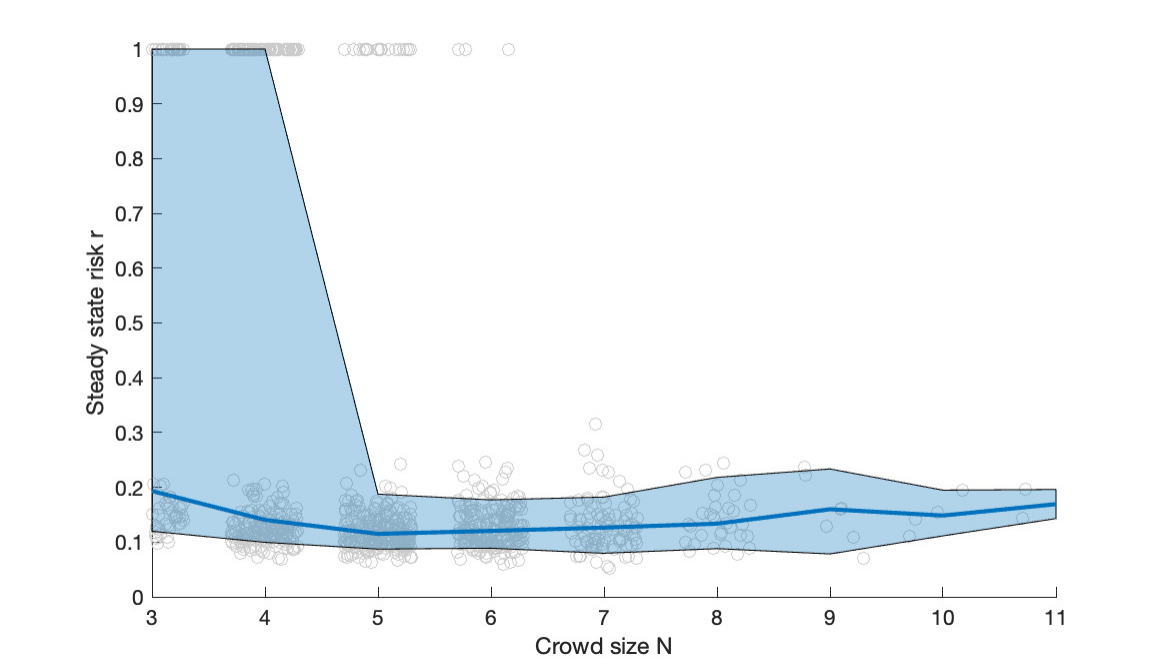}
      \caption{Steady state learned risk in a small world network as a function of the number of witnesses $N$, holding all else constant at the baseline values in Table \ref{tab:param}. Grey dots are individual witness steady state risk levels $r_i$ over the entire network of 1000 nodes, jittered horizontally for easier visualization. The median steady state risk level for each $N$ (blue line) remains approximately constant, but the variance decreases with $N$ (10\% to 90\% quantiles of simulation results for each $N$ shown with blue shading).} \label{fig:network_dynamics}
\end{figure}

In a typical simulation, risk levels for each node settle to a steady state, like in the simpler all-to-all network. Unlike the all-to-all network, learning from neighbors with different degrees leads to substantially different collective behavior, especially for small crowd sizes (see Figure \ref{fig:network_dynamics}). 

For small crowd sizes, the majority of individuals converge on a low risk level, but some individuals converge to the maximum risk level of 1; in contrast, in the all-to-all network with small crowd sizes, the majority of individuals converged to the lowest risk 0, while only a few individuals (if any) converged on a positive low risk level. 

For larger crowd sizes, the collective behavior is qualitatively similar for all-to-all and small world networks. In both situations, individuals converge to small positive risk levels; however, the median steady state risk level is larger for the small world network than for the all-to-all network.

These results show that the all-to-all network convergence to no risk is not robust to a more realistic network structure. In such a network, many individuals exposed to a social network with a distribution of crowd sizes will learn a moderately low risk of action. However, the smaller the crowd size for an individual, the more likely they are to eventually learn that intervening is too risky to ever try (these are the minority of small-$N$ nodes).

That said, a fixed small world network structure is less realistic than the dynamically rewiring social networks seen in the real world, where individuals interact with random numbers of strangers throughout their days and lives.

\subsection{Social Learning on a Continuum} \label{sec:model3}
If we assume that the number of witnesses $N$ is a parameter within a continuum population, this leads to a partial differential equation for the population distribution of risk $r$. Because probability is conserved, the distribution of risk $r$ within a population follows the continuity equation:
\begin{align*}
\frac{\partial p}{\partial t} = - \frac{\partial}{\partial r} \left( p \frac{\mathrm{d}r}{\mathrm{d}t} \right),
\end{align*}
where $p(r,t)$ is the probability density function for $r$, $\ds \frac{\mathrm{d}r}{\mathrm{d}t}$ is given by Equation \eqref{eq:odes} sans index $i$, and the probabilities in the continuum limit become
\begin{align*}
    P^{(I)} &= H\Big(1-(1+x)\big(1-(1-r)^N\big)\Big) \\
    P^{(w)} &= (1-r)^{N-1} \\
    \tilde{P}^{(w)} &= \bigg( P^{(I)} + \big(1-P^{(I)}\big) (1-r) \bigg)^{N-1} - \bigg(\big(1-P^{(I)}\big) (1-r)\bigg)^{N-1} \\
    \tilde{P}^{(p)} &= 1 - \tilde{P}^{(w)} - \big(1-P^{(I)}\big)^{N-1}
\end{align*}
We leave this exploration for future work.

\subsection{Sensitivity Analysis: Technical Details}
The global uncertainty and sensitivity analysis was performed using the methodology outlined in Marino et al. \cite{marino2008methodology}. The base code for the analysis is freely available at the author Denise Kirschner's website \cite{marinoCode}.

In brief, the analysis uses Latin Hypercube Sampling (LHS) of parameter space to simulate uncertainty in model parameters. LHS sampling requires fewer model simulations than simple random sampling without introducing bias \cite{mckay1979comparison}. We used uniform sampling of each parameter about the base values given in Table \ref{tab:param}. The ranges of the uniform samples are available in our code.

We use Partial Rank Correlation Coefficients (PRCCs) to test the sensitivity of model outputs to parameter uncertainty because model outputs depend monotonically on model inputs, but those relationships are not linear trends. As noted by Marino et al. \cite{marino2008methodology}, for linear trends we could have used Pearson correlation coefficient (CC), partial correlation coefficients (PCCs), or standardized regression coefficients (SRC). Had our trends been non-monotonic, we would have used the Sobol method or one of its many extensions \cite{saltelli2002making}.
 
The displayed sensitivities in Figure \ref{fig:sensitivity} are the PRCCs for the model output (median steady state risk $r$) and the model inputs (all model parameters, including the initial condition, but excluding crowd size $N$); we must perform separate sensitivity analyses for each $N$ because the order of the ODE systems varies with $N$. 

As described by Marino et al. \cite{marino2008methodology}, partial rank correlation characterizes the monotonic relationship between inputs $x_j$ and output $y$ after the effects on $y$ of the other inputs are removed. The values of PRCCs fall between $-1$ and $1$, with $1$ indicating the strongest positive rank correlation and $-1$ indicating the strongest negative rank correlation. The significance indicates the probability that the rank correlation is zero (i.e., large significance suggests that there is no relationship between $x_j$ and $y$). 

\end{document}